\documentclass[manuscript]{aastex}
\usepackage{graphicx}
\usepackage{rotating}
\usepackage{chngpage}
\usepackage{array}

\slugcomment{Not to appear in Nonlearned J., 45.}

\shorttitle{Estimating Photometric Redshifts of Quasars}
\shortauthors{Zhang et al.}

\begin{document}

\def \aj {AJ}
\def \mnras {MNRAS}
\def \apj {ApJ}
\def \apjs {ApJS}
\def \apjl {ApJL}
\def \aap {A\&A}
\def \spie {SPIE}


\title{Estimating Photometric Redshifts of Quasars via K-nearest
Neighbor Approach Based on Large Survey Databases  }

\author{Yanxia Zhang\altaffilmark{1},  He Ma\altaffilmark{1}, Nanbo Peng\altaffilmark{1}, Yongheng Zhao\altaffilmark{1}}

\affil{Key Laboratory of Optical Astronomy, National Astronomical
Observatories, Chinese Academy of Sciences 100012, Beijing,
P.R.China} \email{zyx@bao.ac.cn}

\and

\author{Xue-bing Wu\altaffilmark{2}}
\affil{Department of Astronomy, Peking University 100871, Beijing,
P.R.China}

\begin{abstract}
We apply one of lazy learning methods named $k$-nearest neighbor
algorithm ($k$NN) to estimate the photometric redshifts of quasars,
based on various datasets from the Sloan Digital Sky Survey (SDSS),
UKIRT Infrared Deep Sky Survey (UKIDSS) and Wide-field Infrared
Survey Explorer (WISE) (the SDSS sample, the SDSS-UKIDSS sample, the
SDSS-WISE sample and the SDSS-UKIDSS-WISE sample). The influence of
the $k$ value and different input patterns on the performance of
$k$NN is discussed. $k$NN arrives at the best performance when $k$
is different with a special input pattern for a special dataset. The
best result belongs to the SDSS-UKIDSS-WISE sample. The experimental
results show that generally the more information from more bands,
the better performance of photometric redshift estimation with
$k$NN. The results also demonstrate that $k$NN using multiband data
can effectively solve the catastrophic failure of photometric
redshift estimation, which is met by many machine learning methods.
By comparing the performance of various methods for photometric
redshift estimation of quasars, $k$NN based on KD-Tree shows its
superiority with the best accuracy for our case.
\end{abstract}


\section{Introduction}
\label{sec:intro}

Astronomy steps into an era of abundant photometric data with the
development of CCD and other modern technologies. Several wide-field
surveys with large ground- and space-based telescopes will produce
over a hundred million photometric data which are larger than
spectroscopic data by two or three orders of magnitude. The most
representative survey should be the Sloan Digital Sky Survey (SDSS,
York et~al. 2000), which greatly promotes the development of modern
astronomy and improves the methods of analyzing massive astronomical
data. Many advanced data mining and machine learning algorithms (See
reviews written by Zhang et~al. 2008, Borne 2009 and Ball \& Brunner
2010) have been successfully applied to deal with problems caused by
massive astronomical data sets, such as galaxy classification,
quasar candidate selection, photometric redshift estimation and so
on. In the next decade, the ongoing and planned multiband
photometric surveys, for instance, the above mentioned SDSS, the
United Kingdom Infrared Telescope (UKIRT; Lawrence et~al. 2007), the
Large Synoptic Survey Telescope (LSST; Tyson 2002), the Panoramic
Survey Telescope and Rapid Response System (Pan-STARRS; Kaiser
et~al. 2002) and so on, will bring many more challenges for
astronomers. This situation will make the technologies of data
mining and machine learning become more popular among the
astronomical community.

Photometric redshifts are a very important and powerful statistical
tool for providing distances of celestial objects to study the
evolutionary properties of galaxies and cosmology. Photometric
redshifts are those derived from only images or photometry. So far
the techniques of photometric redshifts have been in a bloom while
deep multicolor photometric surveys have been carried out, with lots
of objects inaccessible to spectroscopic observations or too time
consuming with the available instruments. The techniques are divided
into two types: empirical methods and the template-fitting approach.
Empirical methods use a subsample of the photometric data with
spectroscopic redshifts as a training set for the redshift
estimators, this subsample describes the redshift distribution in
magnitude and color space empirically and is utilized to construct
regressors to predict photometric redshifts. A lot of machine
learning techniques show good performance on the problem of
determining photometric redshifts of galaxies or quasars. Artificial
Neural Networks (ANN) have been used by astronomers (Firth et~al.
2003; Collister \& Lahav 2004; Blake et~al. 2007; Oyaizu et~al.
2008; Y$\grave{e}$che et~al. 2010; Zhang et~al. 2009) for
determining photometric redshifts. In recent years, many other
algorithms for photometric redshifts have entered into our sight.
Way et~al. (2009) demonstrated that Gaussian Process Regression
(GPR) can be a competitive approach to ANN and other least-squares
fitting methods. Geach (2011) presented an application of the
Self-Organized Map (SOM) for visualizing, exploring and mining the
catalogues of large astronomical surveys. Weak Gated Experts (WGE)
which allows to derive photometric redshifts through a combination
of data mining techniques was presented by Laurion et~al. (2011).
Ball et~al. (2007; 2008) discussed how to apply a nearest neighbor
algorithm to predict photometric redshifts. Polsterer et~al. (2013)
applied $k$-nearest neighbor approach to estimate photometric
redshifts of quasars and preselect high-redshift quasar candidates.
Wadadekar (2005) employed Support Vector Machines (SVMs) to predict
photometric redshifts of galaxies. However, SVMs is not a good
choice for estimating photometric redshifts (Wang et~al. 2008; Peng
et~al. 2010a). Template methods use either libraries of observed
spectra exterior to the survey or model spectral energy
distributions (SEDs). Since these are full spectra, the templates
can be shifted to any redshift and then convolved with the
transmission curves of the filters used in the photometric survey to
create the template set for the redshift estimators. There are many
works on the SED fitting technique or template fitting technique,
for example, Bolzonella et~al. 2000, Budav$\acute{a}$ri et~al.
(2000, 2001), Babbedge et~al. (2004), Rowan-Robinson et al. (2008),
Wolf et~al. (2008, 2009), Wu et~al. (2010), Ilbert et~al. (2009,
2011), Salvato et~al. (2009, 2011). In addition, some researches
focus on comparison of methods for photometric redshift estimation
(Hildebrandt et~al. 2010; Abdalla et~al. 2011; Oyaizu et~al. 2008;
Wang et~al. 2008).

In this work we attempt to use $k$-nearest neighbor ($k$NN)
algorithm which is a simple algorithm and easy to be understood by
astronomers to estimate photometric redshifts of quasars. This
method will be used to determine the photometric redshifts of
quasars which are much more challenging than those of galaxies
because there is a catastrophic failure in estimating photometric
redshifts of quasars when the spectroscopic redshift is less than 3.
All wide-field quasar photometric redshift results suffer this
problem (e.g. Weinstein et~al. 2004; Richards et~al. 2001; Wu et~al.
2004). In the following we apply $k$-nearest neighbor with multiband
data based on the SDSS (the Sloan Digital Sky Survey), UKIDSS (the
UKIRT Infrared Deep Sky Survey) and WISE (the Wide-field Infrared
Survey Explorer) databases to solve this problem.

This paper is organized as follows. Section~2 describes the
characteristics of data used in this experiment in detail. In
Section~3.1, we present the principle of $k$NN, and discuss some
detail of it. Section~3.2 demonstrates the performance of $k$NN
using multiband datasets for estimating photometric redshifts of
quasars. In Section~4, we give the conclusion and what should be
improved in the future work.

\section{Data}
\label{sec:data}

There are three data sets used in this work. The first set is
generated from the Sloan Digital Sky Survey (SDSS; York et~al. 2000)
which could be one of the most ambitious and influential surveys in
the history of astronomy. Since we focus on the photometric redshift
estimation of quasars, we use the samples of the Quasar Catalogue V
(Schneider et~al. 2010) with highly reliable redshifts as our basic
data set. Based upon the SDSS DR7 (Abazajian et~al. 2009), quasar
Catalogue V contains 105,783 spectroscopically confirmed quasars and
represents the conclusion of the SDSS-I and SDSS-II quasar survey.
We cross-identify all quasars in this table with the UKIRT (UK
Infrared Telescope) Infrared Deep Sky Survey (UKIDSS) (Warren et al.
2000; Lawrence et al. 2007) data release eight (DR8) and the whole
data release of the NASA's Wide-field Infrared Survey Explorer
(WISE; Wright et al. 2010), respectively.

\begin{figure}[!!!h]
\centering
\includegraphics[bb=1 23 386 498,width=6cm,clip]{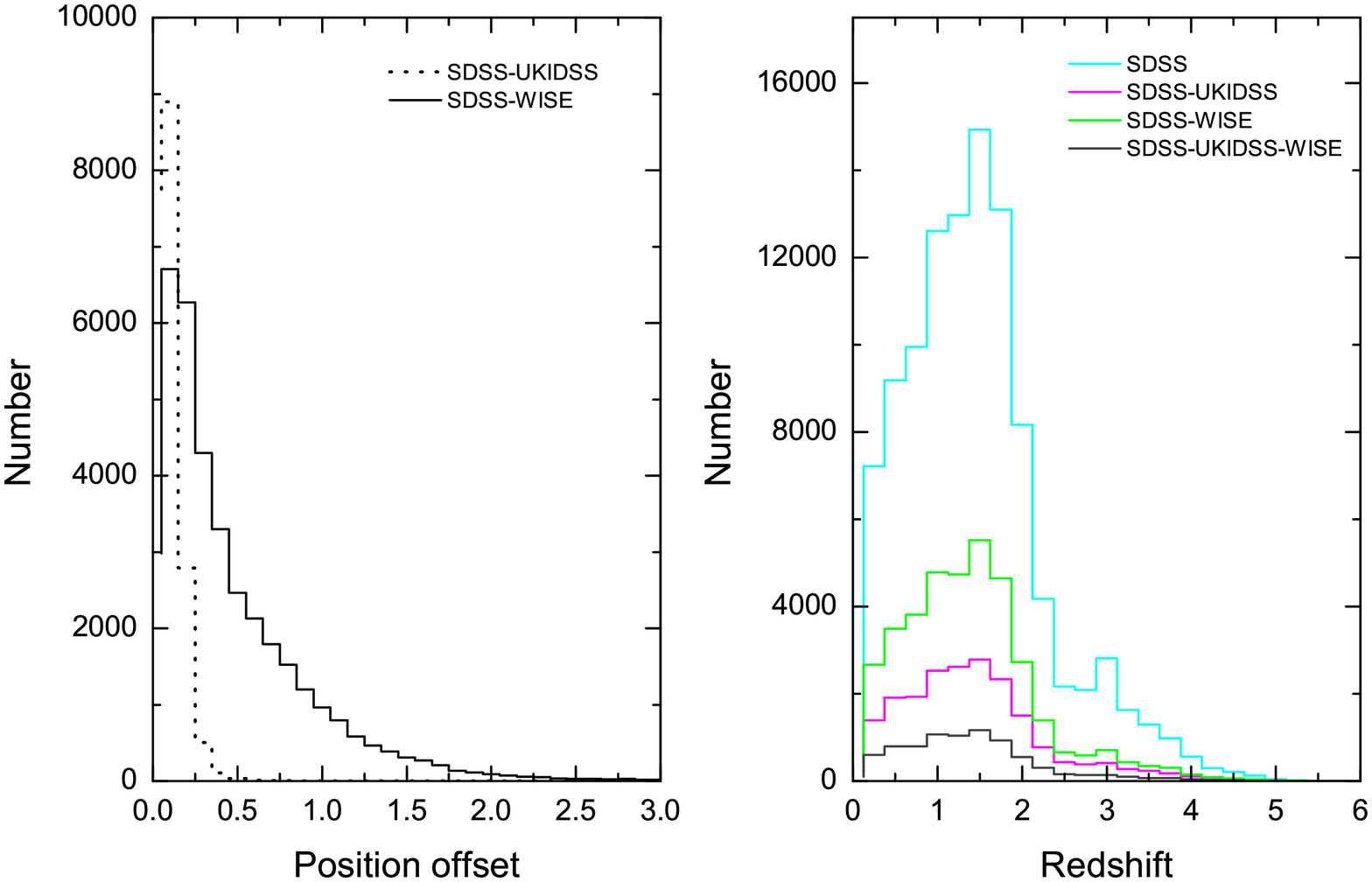}
\includegraphics[bb=1 72 503 691,width=6cm,clip]{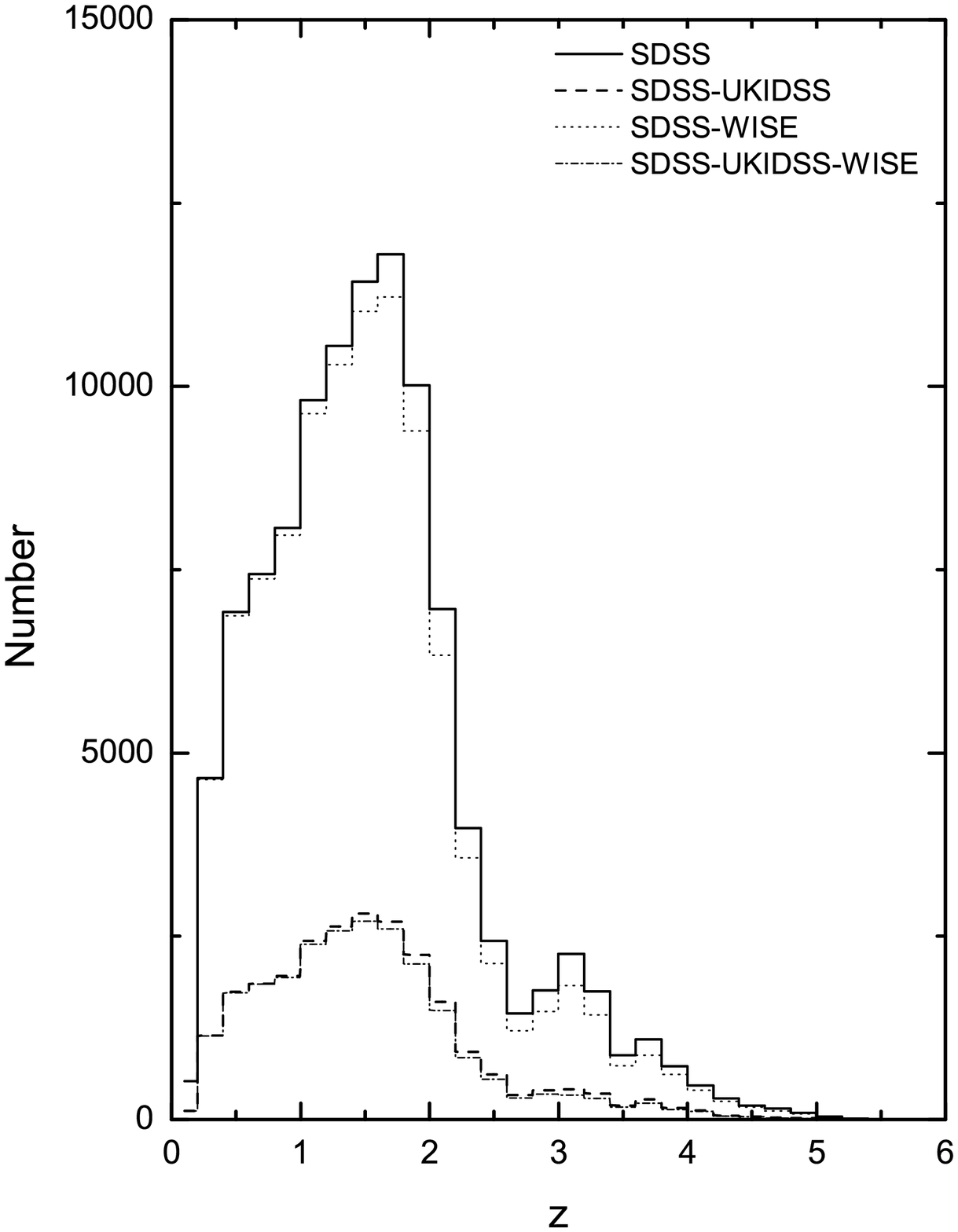}
 \caption{Left: The positional offset distribution of SDSS-UKIDSS
and SDSS-WISE. Right: The quasar redshift distribution of SDSS
(solid line), SDSS-UKIDSS (dash line), SDSS-WISE (dot line) and
SDSS-UKIDSS-WISE samples(dash dot line).} \label{fig:OR}
\end{figure}

For the cross-matching of Quasar Catalogue V with UKIDSS DR7, we
adopt the closest counterparts within 3 arcsec radius between the
positions in the two catalogs. In order to avoid the area of lower
Galactic latitudes which is crowed with many stellar objects, we
only adopt the data in the UKIDSS Large Area Survey (LAS) (Wu et~al.
2010). In the left panel of Figure~\ref{fig:OR}, the distribution of
the positional offset between SDSS and UKIDSS is shown by a dotted
line. It clearly shows that almost all (99.68\%) of the closest
counterparts are within 0.5 arcsec of the SDSS positions. We then
cross-match Quasar Catalogue V with WISE pre within 6 arcsec radius
given that the angular resolution of WISE is 6.1 arcsec, 6.4 arcsec,
6.5 arcsec and 12.0 arcsec in the four bands (W1, W2, W3, W4),
respectively. In the left panel of Figure~1, the positional offset
distribution of the cross-match result of the two data sets is
indicated by a solid line and it is found that 86.74\% of the
cross-match sources have offsets smaller than 1 arcsec and 95.57\%
of them have offsets smaller than 2 arcsec.

As a result, in order to improve the probability and reliability of
cross-identification, the cross-match radius between SDSS and UKIDSS
DR8 is taken 0.5 arcsec, the radius between SDSS and WISE is set 2
arcsec. Thus the number of quasars for the SDSS-UKIDSS and SDSS-WISE
samples is 25184 and 100208, respectively. For the SDSS-UKIDSS and
SDSS-WISE samples obtained by the front steps, the same SDSS ID of
sources are collected as the SDSS-UKIDSS-WISE sample. The number of
this sample is 24089. Consequently, the number of SDSS-UKIDSS-WISE
quasars is smaller than the other two quasar samples.

In the right panel of Figure~1, we plot the redshift histogram of
quasars for the four samples of SDSS, SDSS-UKIDSS, SDSS-WISE and
SDSS-UKIDSS-WISE. The distributions of the SDSS and SDSS-WISE
samples are very similar, those of the SDSS-UKIDSS and
SDSS-UKIDSS-WISE samples are just alike. Considering Galactic
extinction, magnitudes from SDSS are corrected by the dust reddening
map of Schlegel et~al. (1998). Since UKIDSS and WISE surveys both
use the Vega Magnitude system, we decide to convert dereddened SDSS
AB magnitudes to Vega magnitudes using the rules described by Hewett
et~al. (2006). In the following experiments, $u,g,r,i,z$ all points
to dereddened SDSS $u,g,r,i,z$ in Vega system, respectively.

\section{METHOD}
\label{sec:method}
\subsection{$k$NN}

In pattern recognition, the $k$-nearest neighbor algorithm ($k$NN)
is a simple method for classifying objects by the closest training
examples in the feature space when data labels are discrete or for
regressing objects based on the average of its nearest neighbors in
the feature space when data labels are continuous. The solution is
defined as follows: given a collection of data points and a query
point in an $m$-dimensional feature space, find the data point
closest to the query point. It is a type of instance-based learning,
or lazy learning algorithm (no explicit training step is required)
that has been shown to be very effective for a variety of problems
in many fields. When $k$NN is used to estimate photometric
redshifts, it simply assigns the property value of the unknown
object to be the average of the spectral redshift values of its k
nearest neighbors. Sometimes it is helpful to weight points so that
adjacent points contribute more to the predicted value than distant
points. The main idea behind $k$NN is that similar training samples
have similar output values, but it is sensitive to the local
structure of the data.

In the field of machine learning, fast nearest neighbor search is a
hot issue. Given the high computational cost of $k$NN, the modern
$k$NN normally works based on some data structure such as KD-Tree or
Ball-Tree. When the dimensionality of sample is less than 20,
KD-Tree is a better choice for $k$NN. Ball-Tree is more suitable for
high-dimensional problems and the brute force approach can be more
efficient than a tree-based approach when dealing with small
samples. For brute-force $k$NN, the distances between all pairs of
points in the sample need to be computed during the neighbor search.
Apparently, brute-force neighbor searches are very appropriate for
small samples. But while the number of samples increases, the
brute-force approach can't work quickly. To overcome this
difficulty, various tree-based data structures have been proposed.
The KD-Tree data structure holds the points in a tree, which
recursively splits the points into two groups along data axes and
stops until one of the termination criteria is met. It is very fast
to construct a KD-Tree. Although the KD-Tree approach shows great
efficiency in low-dimensional neighbors search, it becomes invalid
while the dimension becomes very large due to so called ``curse of
dimensionality". Based on this status, the Ball-Tree data structure
was designed to enable fast nearest neighbor searching in
high-dimensional spaces. The Ball-Tree is made up of a series of
``balls" and ``nodes". In the Ball-Tree structure, the internal
node, a node within a node is differentiated by the region including
all of its derivative balls. Based on the spherical geometry of the
Ball-Tree nodes, such trees might perform faster if they capture the
true distribution of data in a high dimensional space. About the
detailed introduction and comparison of brute force, KD-Tree and
Ball-Tree can refer to the web \\``
http://scikit-learn.org/stable/modules/neighbors.html".

For estimating photometric redshifts of quasars with multiband
datasets, kNN based on KD-Tree should be the right choice. There is
a small detail of $k$NN needed to be discussed that which kind of
distance is the best measured method for determining photometric
redshifts. Actually, we have compared the performances of three
kinds of distances (Euclidean, Chebychev and Cityblock) used by
$k$NN for quasar candidate selection (Peng et al. 2010b). Euclidean
distance is the most common use of distance. When people talk about
distance, this is what they are referring to. Euclidean distance
examines the root of square differences between the coordinates of a
pair of objects. Chebyshev distance (i.e. the Maximum value
distance) defined on a vector space where the distance between two
vectors is the greatest of their differences along any coordinate
dimension. In other words, it examines the absolute magnitude of the
differences between the coordinates of a pair of objects. Cityblock
distance, also named Manhattan distance, examines the absolute
differences between the coordinates of a pair of objects. The
performance of $k$NN has no obvious difference using different
distance definition. We even try to use Cosine distance to improve
the performance of $k$NN for photometric redshift estimation but
there is no sufficient evidence to prove which kind of distance is
best suitable for our problem. Therefore, in this work we use
Euclidean distance to find the $k$-nearest neighbor of the unknown
object for determining photometric redshifts.

When using multiband data to estimate photometric redshifts of
quasars, the input parameters of an algorithm are various
magnitudes, colors and spectral redshifts. The algorithms try to
find a mapping relationship between magnitudes, colors and spectral
redshifts. Then the redshifts of unknown objects can be measured by
this relationship. For estimating photometric redshifts of quasars,
there are two important evaluation indexes: the percents in
different $|\Delta z| $ ($|\Delta z|=|\frac{z_{\rm phot}-z_{\rm
spec}}{1+z_{\rm spec}}|$) intervals and the Root Mean Square ($RMS$)
of $|\Delta z|$ which can be considered as accuracy and dispersion
of predicted redshifts of $k$NN, respectively. Since the photometric
redshift estimation of quasars is much more difficult than that of
galaxies, we not only consider the value of $RMS$ which is a
statistical measure of the magnitude of a varying quantity, but also
compute its accuracies in three $|\Delta z|$ regions in order to
know the overall performance of a certain algorithm. In the
following experiments, we make use of the two evaluation indexes to
compare the performances of $k$NN for determining photometric
redshifts of quasars with various samples.

\subsection{Results and discussions}

In this section, we compare the performances of $k$NN using the
samples of SDSS, SDSS-UKIDSS, SDSS-WISE and SDSS-UKIDSS-WISE with
different input patterns, respectively. In addition, we separately
analyze the SDSS-UKIDSS, SDSS-WISE and SDSS-UKIDSS-WISE samples,
only using SDSS features just like that has been done by Ball et~al.
( 2007; 2008).

Considering poor performance of estimating photometric redshifts of
quasars only using SDSS optical band data, we utilize multiband data
from optical to infrared band to improve it. Many authors (Ball
et~al. 2007; Bovy et~al. 2011; Way et al. 2009; Wu et~al. 2010) have
successfully used multiband data to get better performance of
photometric redshift estimation of galaxies or quasars. In this
Subsection, the four datasets (the SDSS sample, the SDSS-UKIDSS
sample, the SDSS-WISE sample and the SDSS-UKIDSS-WISE sample) from
three sky surveys are used to evaluate the performance of $k$NN for
phorometric redshift estimation of quasars. When $k$NN are employed
for photometric redshift estimation, the known samples are randomly
divided into a third and two-thirds for ten times, two-thirds for
training and a third for testing. Therefore the experiment performs
ten times for each input pattern.

There are various parameter combination from different datasets.
Which is the best input pattern? This is a difficult question to
answer for it is different for different samples with different
algorithms. Moreover, it has different model parameters to adjust
for the same algorithm. Only when the algorithm and its model
parameters as well as the test data are set is the best input
pattern fixed. As a result, the $k$ value of $k$NN is firstly set as
1 in the following experiments. Thus it is easy to find good input
pattern for the same data and algorithm. In the second step, we try
to find the best $k$ value when the input pattern is set for a
special data sample.

Firstly, we investigate the solution of determining photometric
redshifts with just the SDSS optical band. In the dissertation of
Kumar (2008), he used Artificial Neural Networks (ANNs) to estimate
photometric redshifts with several different SDSS input patterns
(parameter combinations). From those experiments described in that
dissertation, it showed that the performance using five $ugriz$
magnitudes is very similar to that of four $u-g,g-r,r-i,i-z$ colors
($u-g,g-r,r-i,i-z$, hereafter short for $4C$). If these features
were combined together, there is a very little improvement on the
performance of ANNz for photometric redshift estimation of quasars.
In Table 1, we redo these experiments with $k$NN and get the same
conclusions. In addition, the pattern of the four colors with $r$ or
$i$ magnitude is tried. The experimental result indicates that the
performance of $k$NN increases when the input pattern $4C+r$ or
$4C+i$ is considered, compared to the results with $4C$, $5Mag$ or
$4C+5Mag$. The performance with $4C+r$ is a little better than that
with $4C+i$.

\begin{table}
\begin{center}
\caption{Comparison of the performances of the nearest neighbor
method ($k=1$) on different samples and different input patterns.
$5Mag = u, g, r, i, z$; $4C = u-g,g-r,r-i,i-z$; $8C =
u-g,g-r,r-i,i-z,z-Y,Y-J,J-H,H-K$; $6C= u-g,g-r,r-i,i-z,z-W1,W1-W2$;
$8C' = u-g,g-r,r-i,i-z,z-W1,W1-W2,W2-W3,W3-W4$; $10C = u-g, g-r,
r-i, i-z, z-Y, Y-J, J-H, H-K, K-W1, W1-W2$; $12C = u-g, g-r, r-i,
i-z, z-Y, Y-J, J-H, H-K, K-W1, W1-W2, W2-W3, W3-W4$.
}\label{tab:perf} {\tiny

\begin{tabular}{lccccc}
\hline \hline
Data Set & Input Pattern & $|\Delta z| <0.1$ (\%) & $|\Delta z| <0.2$ (\%) & $|\Delta z| <0.3$ (\%) & $RMS$ \\
\hline
SDSS & $5Mag $ & 76.78 $\pm$ 0.31 & 84.61 $\pm$ 0.21 & 86.18 $\pm$ 0.16 & 0.272 $\pm$ 0.002 \\
SDSS & $4C $ & 76.72 $\pm$ 0.34 & 83.67 $\pm$ 0.33 & 85.19 $\pm$ 0.32 & 0.287 $\pm$ 0.004 \\
SDSS & $4C,5Mag$ & 78.07 $\pm$ 0.23 & 85.39 $\pm$ 0.20 & 86.80 $\pm$ 0.33& 0.263 $\pm$ 0.003\\
SDSS & $4C,r$ & \textbf{78.63 $\pm$0.23} &\textbf{85.70 $\pm$ 0.27} & \textbf{ 87.09 $\pm$ 0.23 }& \textbf{0.259$\pm$ 0.003} \\
SDSS & $4C,i$ & 78.59 $\pm$ 0.25 & 85.65$\pm$ 0.31 & 87.05 $\pm$ 0.36& 0.260 $\pm$ 0.002\\
SDSS-UKIDSS & $4C $ & 77.00 $\pm$ 0.70 & 83.88 $\pm$ 0.52 & 85.43 $\pm$ 0.59 & 0.285 $\pm$ 0.007 \\
SDSS-UKIDSS & $4C,Y-J$ & 87.78 $\pm$ 0.61 & 88.97 $\pm$ 0.40 & 90.33 $\pm$ 0.50 & 0.232 $\pm$ 0.004 \\
SDSS-UKIDSS & $4C,Y-J,Y-K $ & 87.71 $\pm$ 0.50 & 93.12 $\pm$ 0.50 & 94.55 $\pm$ 0.46 & 0.174 $\pm$ 0.013 \\
SDSS-UKIDSS & $8C$ & 91.16 $\pm$ 0.38 & 94.72 $\pm$ 0.54 & 95.85 $\pm$ 0.43 & 0.150 $\pm$ 0.006 \\
SDSS-UKIDSS & $8C,r$ & \textbf{91.18 $\pm$ 0.46 }& \textbf{94.98 $\pm$ 0.50} & \textbf{96.04 $\pm$ 0.27} & \textbf{0.148 $\pm$ 0.008} \\
SDSS-UKIDSS & $8C,i$ & 90.99 $\pm$ 0.37& 94.91 $\pm$ 0.33 & 95.99 $\pm$ 0.26 & 0.148 $\pm$ 0.008 \\
SDSS-WISE & $4C$ & 76.80 $\pm$ 0.32 & 87.73 $\pm$ 0.23 & 85.20 $\pm$ 0.40 & 0.286 $\pm$ 0.003 \\
SDSS-WISE & $6C$ &87.96 $\pm$ 0.25 & 95.25 $\pm$ 0.16 & 96.78 $\pm$ 0.14 & 0.132 $\pm$ 0.003 \\
SDSS-WISE & $8C'$ & 86.83 $\pm$ 0.29 & 95.53 $\pm$ 0.25 & 97.20 $\pm$ 0.13& 0.127 $\pm$ 0.003 \\
SDSS-WISE & $8C',r$ & 86.19$\pm$ 0.19 &95.86 $\pm$ 0.17 & 97.52 $\pm$ 0.13& 0.120 $\pm$ 0.0003 \\
SDSS-WISE & $6C,r$ & 88.56 $\pm$ 0.27&  96.36 $\pm$ 0.23 & 97.63 $\pm$ 0.11 & 0.117 $\pm$ 0.005 \\
SDSS-WISE & $6C,i$ & \textbf{88.64 $\pm$ 0.25} &  \textbf{96.38 $\pm$ 0.15} & \textbf{97.63 $\pm$ 0.10} & \textbf{0.117 $\pm$ 0.004} \\
SDSS-UKIDSS-WISE & $ 4C $ & 76.60 $\pm$ 0.82 & 83.61 $\pm$ 0.83 & 85.20 $\pm$ 0.77 & 0.284 $\pm$ 0.009 \\
SDSS-UKIDSS-WISE & $ 8C $ & 91.44 $\pm$ 0.50 & 94.90 $\pm$ 0.47 & 95.96 $\pm$ 0.44 & 0.150 $\pm$ 0.012 \\
SDSS-UKIDSS-WISE & $ 8C' $ & 83.95 $\pm$ 0.91 & 94.40 $\pm$ 0.26 & 96.61 $\pm$ 0.26 & 0.139 $\pm$ 0.011 \\
SDSS-UKIDSS-WISE & $ 12C $ & 91.13 $\pm$ 0.49 & 96.57 $\pm$ 0.29 & 97.90 $\pm$ 0.24 & 0.110 $\pm$ 0.006 \\
SDSS-UKIDSS-WISE & $ 10C $ & \textbf{92.96 $\pm$ 0.33}& 96.89 $\pm$ 0.21 & 98.00 $\pm$ 0.16 & 0.104 $\pm$ 0.008 \\
SDSS-UKIDSS-WISE & $ 10C, r $ & 92.75 $\pm$ 0.63 & 97.21 $\pm$ 0.22 & 98.21 $\pm$ 0.27 & 0.099 $\pm$ 0.011 \\
SDSS-UKIDSS-WISE & $ 10C, i $ & 92.93 $\pm$ 0.29 & \textbf{97.26 $\pm$ 0.36} & \textbf{98.23 $\pm$ 0.09} & \textbf{0.099 $\pm$ 0.009} \\

\hline
\end{tabular}}
\end{center}
\end{table}

For the SDSS-UKIDSS data set, six input patterns are tested by
$k$NN. In order to know the veritable improvement comes from the
addition of UKIDSS infrared bands, we use $k$NN with the input
pattern $4C$ to estimate photometric redshifts of quasars in this
data set. Actually, we do the same test in the data sets (the
SDSS-WISE and SDSS-UKIDSS-WISE samples). From Table 1, it is found
that the influence of the sample size on the performance of $k$NN is
very small, comparing the performances of the SDSS, SDSS-UKIDSS,
SDSS-WISE and SDSS-UKIDSS-WISE samples with the same pattern $4C$.
Considering more information from more bands, the performance of
$k$NN usually rises even without adding parameters. This result
provides us information that it is fair for evaluating the
performance of $k$NN with multiband datasets. The accuracy within
$|\Delta z|<0.3$ and $RMS$ of $k$NN using the input pattern $4C$ is
about $85.43\% \pm 0.59\%$ and $0.285 \pm 0.007$, respectively. We
also test $k$NN with three kinds of parameter combinations ($4C,
Y-J$; $4C, Y-J, Y-K$; $4C, z-Y, Y-J, J-H, H-K$). From Table~1, it
shows that $k$NN can effectively use the additional information from
multiband data. As the information increases, the performance of
$k$NN improves. Adding $r$ or $i$ magnitude to the pattern $8C$ ($8C
= 4C, z-Y, Y-J, J-H, H-K$), the performance of $k$NN further
ascends. The best result is produced by using the input pattern $8C,
r$ and the accuracy during $|\Delta z|<0.3$ reaches $96.04\% \pm
0.27\%$ and $RMS$ amounts to $0.148 \pm 0.008$. This is a great
improvement for estimating photometric redshifts of quasars. The
accuracy within $|\Delta z|<0.1$ using this input pattern reaches
$91.18\% \pm 0.46\%$ and this makes the result produced by this
method becomes more practical.

For the SDSS-WISE data set, we firstly evaluate the performance of
$k$NN by adding WISE W1 and W2 to the SDSS optical band considering
the larger uncertainty of W3 and W4. The result shows that the
accuracy with this input pattern $6C$ ($6C = 4C, z-W1, z-W2$) can
reach $87.96\% \pm 0.25\%$ when $|\Delta z|<0.1$, $95.25\% \pm
0.16\%$ when $|\Delta z|<0.2$, $96.78\% \pm 0.14\%$ when $|\Delta
z|<0.3$ and $RMS$ is $0.132 \pm 0.003$. When further considering
$W2-W3, W3-W4$, the accuracies of $|\Delta z|<0.2$, $|\Delta z|<0.3$
and $RMS$ all increase, but that of $|\Delta z|<0.1$ decreases,
comparing to the pattern $6C$. Continuously adding $r$ magnitude,
i.e. the pattern $4C, z-W1, W1-W2, W2-W3, W3-W4, r$, the all
accuracy indexes except that of $|\Delta z|<0.1$ enhance. In
addition, the pattern of $6C$ adding $r$ or $i$ magnitude is tested.
Table~1 indicates that the performance with $6C+r$ or $6C+i$ is
better than other patterns for this SDSS-WISE sample. The best
pattern is $6C+i$, the accuracy of $|\Delta z|<0.1$, $|\Delta
z|<0.2$ and $|\Delta z|<0.3$ is $88.64\% \pm 0.25\%$, $96.38\% \pm
0.15\%$ and $97.63\% \pm 0.10\%$, respectively, the value of $RMS$
amounts to $0.117 \pm 0.004$. Compared with that of $k$NN using the
SDSS-UKIDSS data set, the performance of $k$NN increases on $RMS$
and the accuracies of $|\Delta z|<0.2$ and $|\Delta z|<0.3$, but
decreases especially for $|\Delta z|<0.1$. We add the four bands of
WISE to the SDSS optical band and want to confirm whether the more
bands can improve the performance of $k$NN that happened in the
experiment of the SDSS-UKIDSS data set. In Table~1, it demonstrates
that $RMS$ of $k$NN decreases and the accuracy improves comparing
only with the SDSS optical information; the accuracies within
$|\Delta z|<0.2$ and $|\Delta z|<0.3$ and $RMS$ are better and the
accuracy within $|\Delta z|<0.1$ is worse than those of the
SDSS-UKIDSS data. Apparently, adding WISE bands can effectively
reduce the $RMS$ value but improve the accuracy within $|\Delta
z|<0.1$ limitedly.

For the SDSS-UKIDSS-WISE data set, we combine all information in the
three surveys to estimate photometric redshifts of quasars. At the
beginning, we still use the $4C$ input pattern to test because this
number of cross-match dataset is relative small (24089). Table~1
indicates that the performances of $k$NN are a little different
between the SDSS-UKIDSS-WISE data set and the SDSS data set with the
same input pattern (4C). For the SDSS-UKIDSS and SDSS-UKIDSS-WISE
samples, the performances with $8C$ as input pattern are similar.
The accuracies within three $|\Delta z|$ ranges and $RMS$ with $8C'$
($8C' = 4C, z-W1, W1-W2, W2-W3, W3-W4$) as input pattern are better
than those with $4C$; the accuracy within $|\Delta z|<0.3$ and $RMS$
have a little improvement comparing to those with the input pattern
$8C$ ($8C = 4C, z-Y, Y-J, J-K, K-H$) while the accuracies within
$|\Delta z|<0.1$ and $|\Delta z|<0.2$ fall down. Now, we use all
information in the three surveys to make a parameter combination
with twelve colors ($12C = u-g, g-r, r-i, i-z, z-Y, Y-J, J-H, H-K,
K-W1, W1-W2, W2-W3, W3-W4$) and this input pattern arrives at a good
result. As Table~1 shown, the overall performance of $k$NN improves
with this input pattern but the accuracy of $|\Delta z|<0.1$ is not
good enough. Therefore, we decide to remove two bands WISE W3 and W4
which are not enough accurate to improve the performance of $k$NN
during $|\Delta z|<0.1$. Really, the two performance indexes all
improve with the input pattern $10C$ ($10C = u-g, g-r, r-i, i-z,
z-Y, Y-J, J-H, H-K, K-W1, W1-W2$). Considering $r$ or $i$ magnitude
in this input pattern $10C$, the accuracies within $|\Delta z|<0.2$
and $|\Delta z|<0.3$ increase and $RMS$ decreases, while the
accuracy within $|\Delta z|<0.1$ tunes smaller comparing to those
with $10C$. Only given the accuracies within $|\Delta z|<0.2$ and
$|\Delta z|<0.3$ as well as $RMS$, the best result is obtained with
$10C+i$ for $k$NN with the accuracy of $97.26\% \pm 0.36\%$ for
$|\Delta z|<0.2$, $98.23\% \pm 0.09\%$ for $|\Delta z|<0.3$ and
$RMS$ of $0.099 \pm 0.009$. This also proves that the precision of
magnitudes influences the performance of $k$NN to predict
photometric redshifts of quasars and the performance of $k$NN maybe
increase or decrease when adding parameters with large uncertainty.

As Table~1 shown, the comparisons of photometric redshift estimation
results are presented for the SDSS sample with the best input
pattern of $4C+r$, the better input pattern of $4C+i$; for the
SDSS-UKIDSS sample with the best input pattern of $8C+r$, the better
input pattern $8C+i$; for the SDSS-WISE sample with the best input
pattern of $6C+i$, the better input pattern of $6C+r$; for the
SDSS-UKIDSS-WISE sample with the best input pattern of $10C+i$, the
better pattern of $10C+r$. All the results are obtained by $k$NN
with $k=1$. Since the better input patterns are determined for the
four datasets, we want to check which value of $k$ is set when the
performance of $k$NN achieves the best for a special input pattern.
For this goal, how the accuracies within the three $|\Delta z|$
ranges and $RMS$ change with different $k$ is shown in Figure~2. In
the left panels of Figure~2, the first plot corresponds to the
pattern $4C+r$ for the SDSS sample, the second to the pattern $8C+r$
for the SDSS-UKIDSS sample, the third to the pattern $6C+r$ for the
SDSS-WISE sample, the fourth to the pattern $10C+r$ for the
SDSS-UKIDSS-WISE sample; in the right panels of Figure~2, the first
plot corresponds to the pattern $4C+i$ for the SDSS sample, the
second to the pattern $8C+i$ for the SDSS-UKIDSS sample, the third
to the pattern $6C+i$ for the SDSS-WISE sample, the fourth to the
pattern $10C+i$ for the SDSS-UKIDSS-WISE sample. In each plot, the
trend of $RMS$ is similar, firstly descends, then ascends. For most
patterns, the three accuracies begin to rise, then drop. While for
the SDSS sample and the SDSS-UKIDSS sample, the accuracies within
$|\Delta z|<0.1$ decline all along. For brief, the appropriate value
of $k$ neighbors is listed in Table~2 for different samples with the
better patterns, separately. Comparing the result in Table~2 with
that in Table~1, the accuracy within $|\Delta z|<0.3$ improves and
$RMS$ decreases while the accuracies within $|\Delta z|<0.2$ and
$|\Delta z|<0.3$ drop for the SDSS and SDSS-UKIDSS samples; the
three accuracies increase and $RMS$ falls for the SDSS-WISE and
SDSS-UKIDSS-WISE samples.

\begin{figure}[!!!h]
\includegraphics[bb=28 180 594 586,width=6cm,clip]{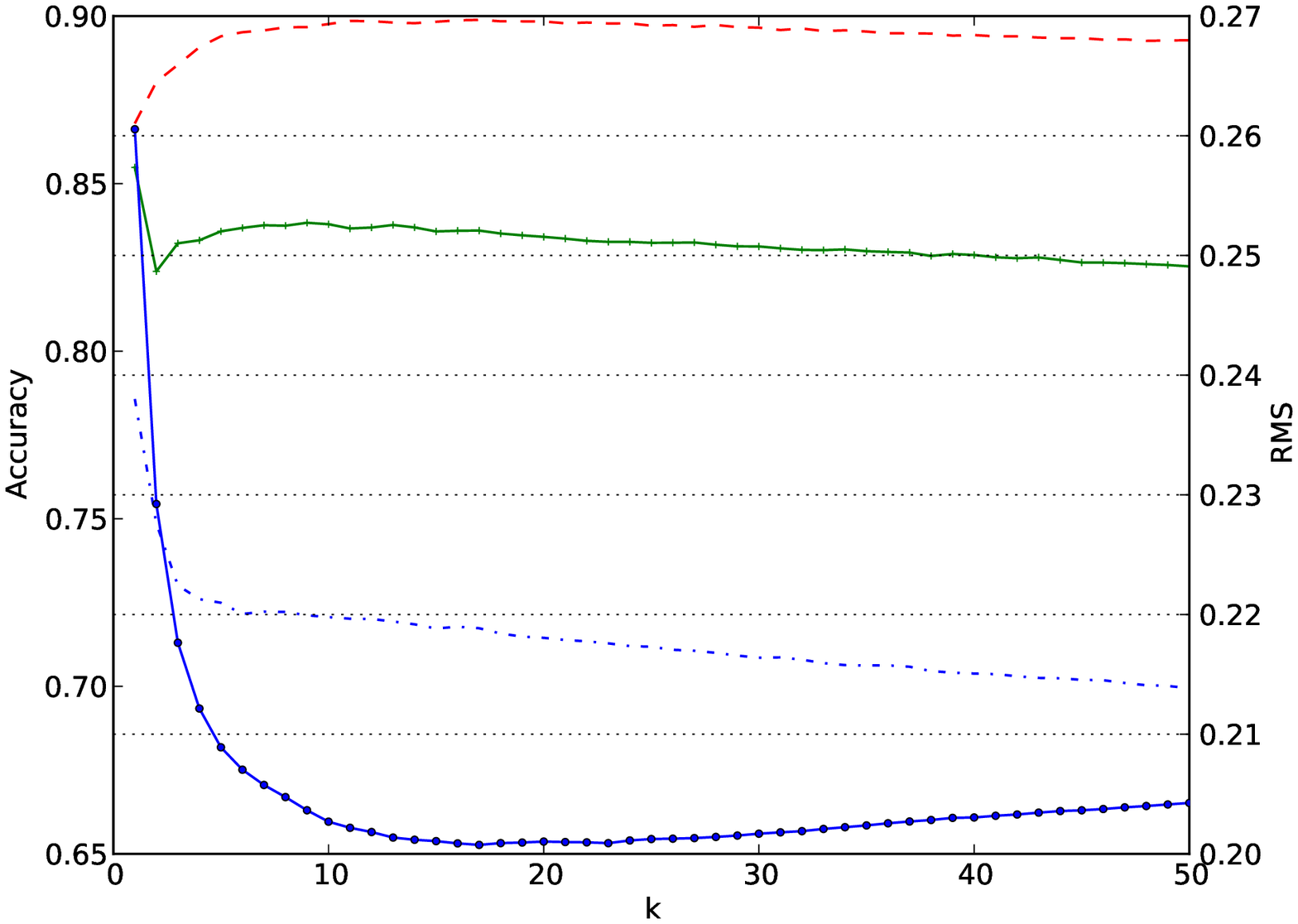}
\includegraphics[bb=28 180 594 586,width=6cm,clip]{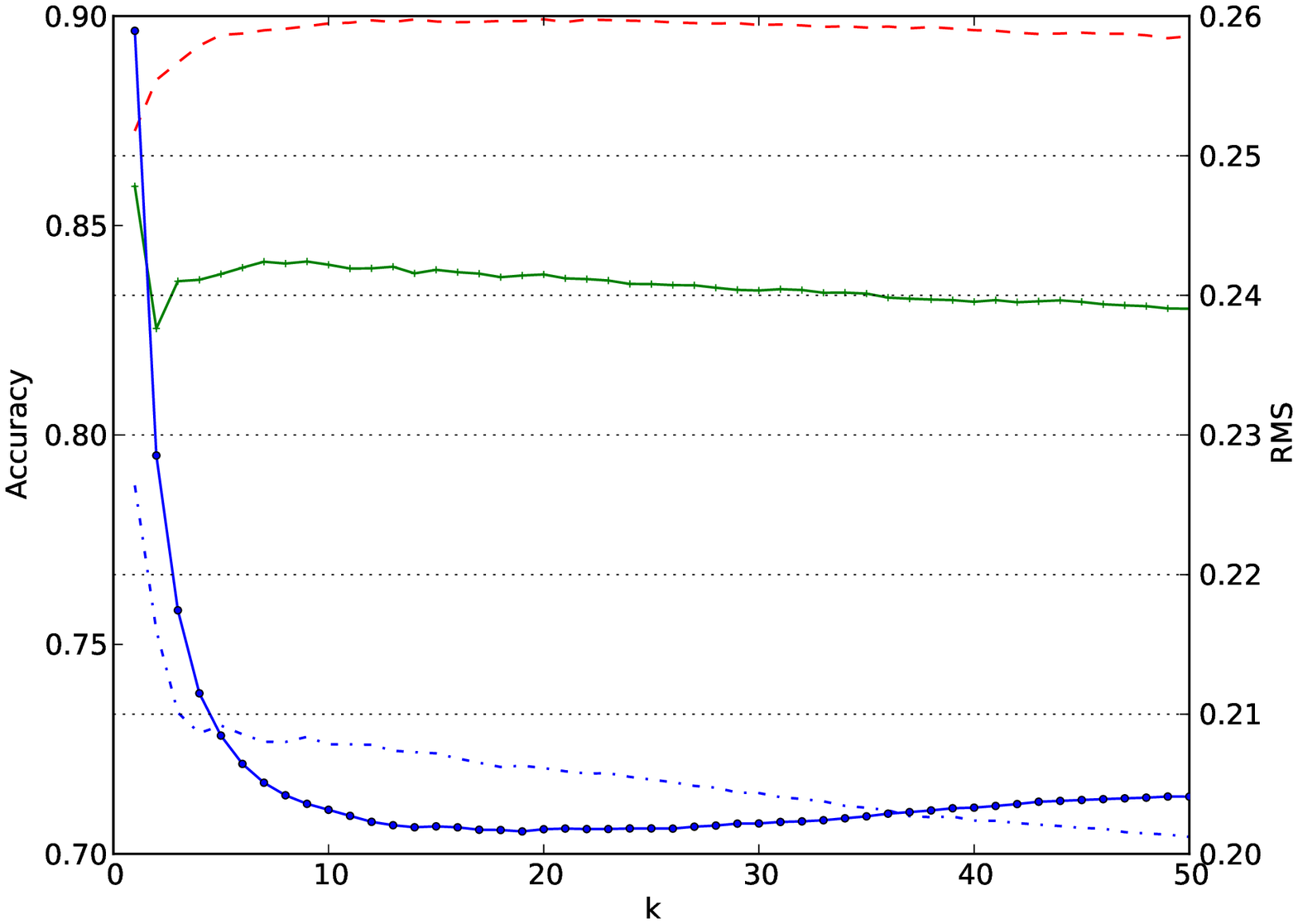}
\includegraphics[bb=28 180 594 586,width=6cm,clip]{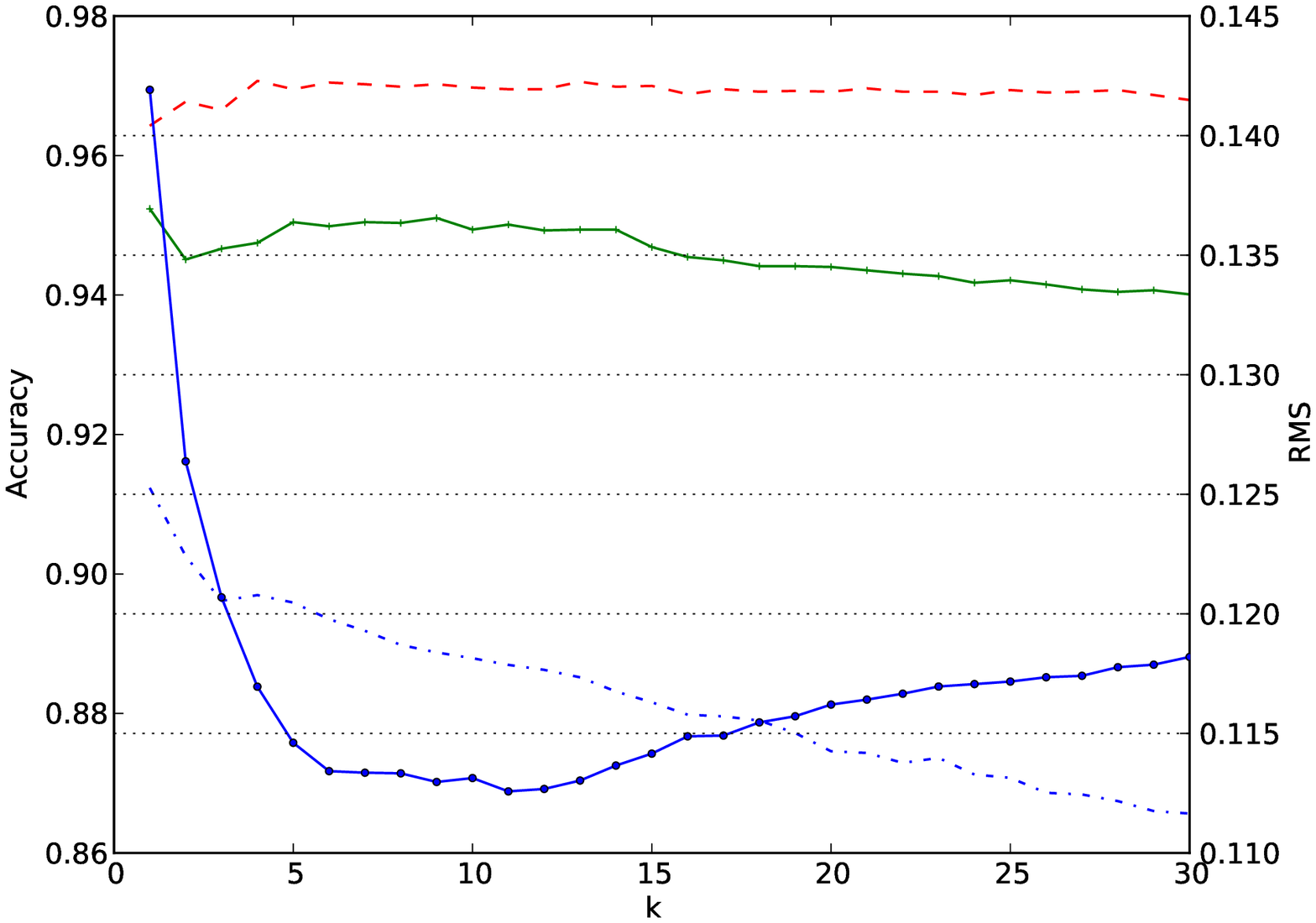}
\includegraphics[bb=28 180 594 586,width=6cm,clip]{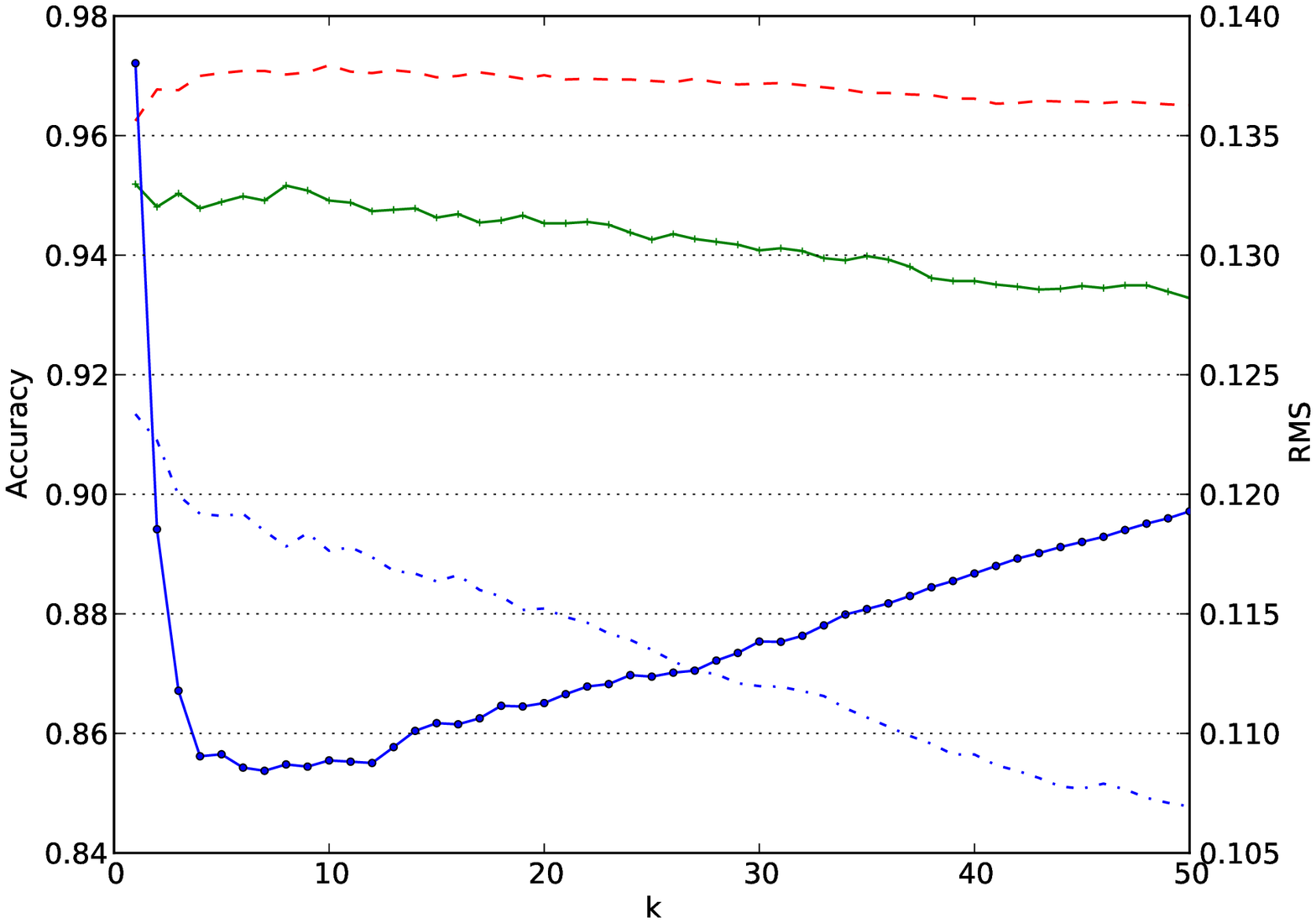}
\includegraphics[bb=28 180 594 586,width=6cm,clip]{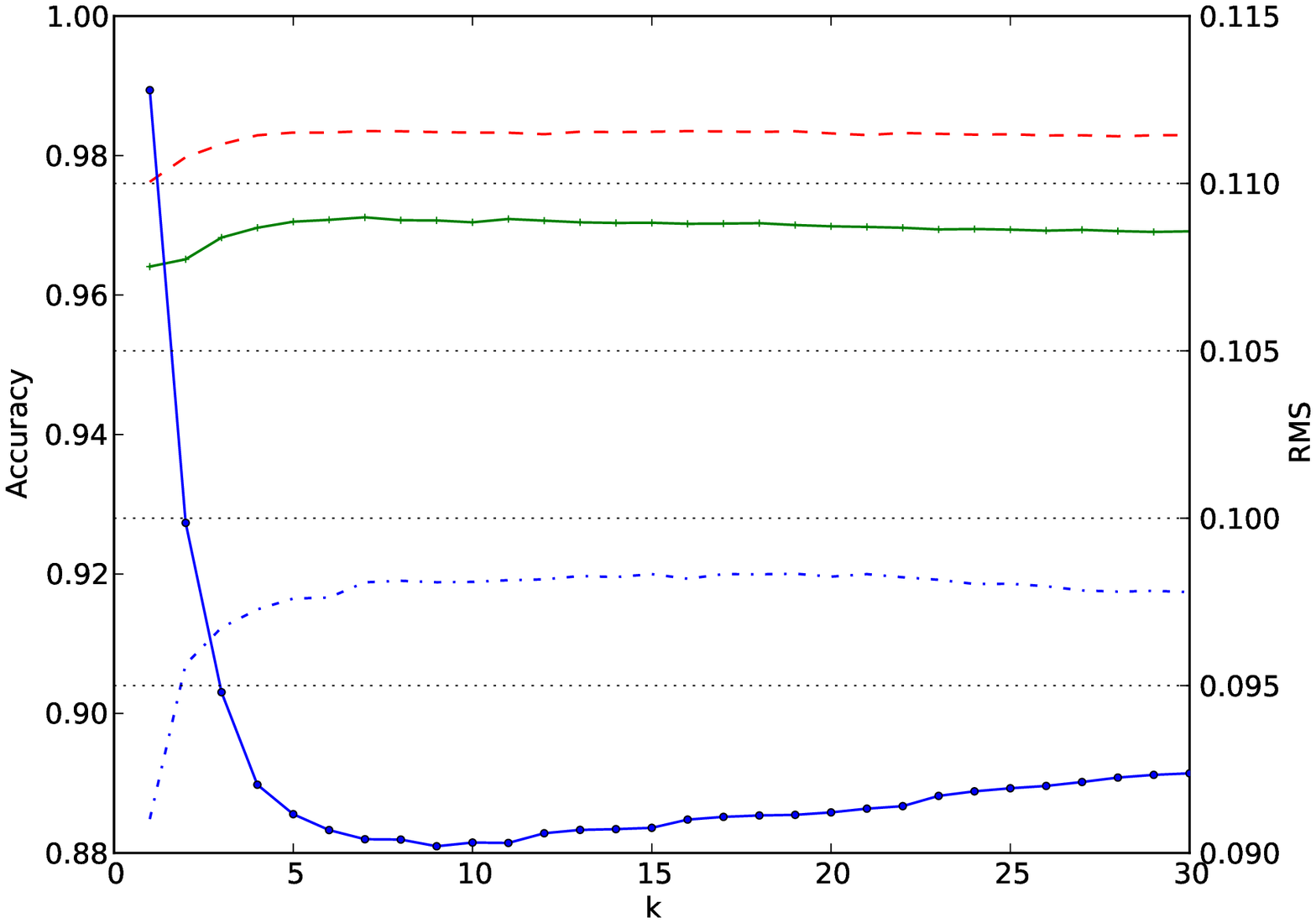}
\includegraphics[bb=28 180 594 586,width=6cm,clip]{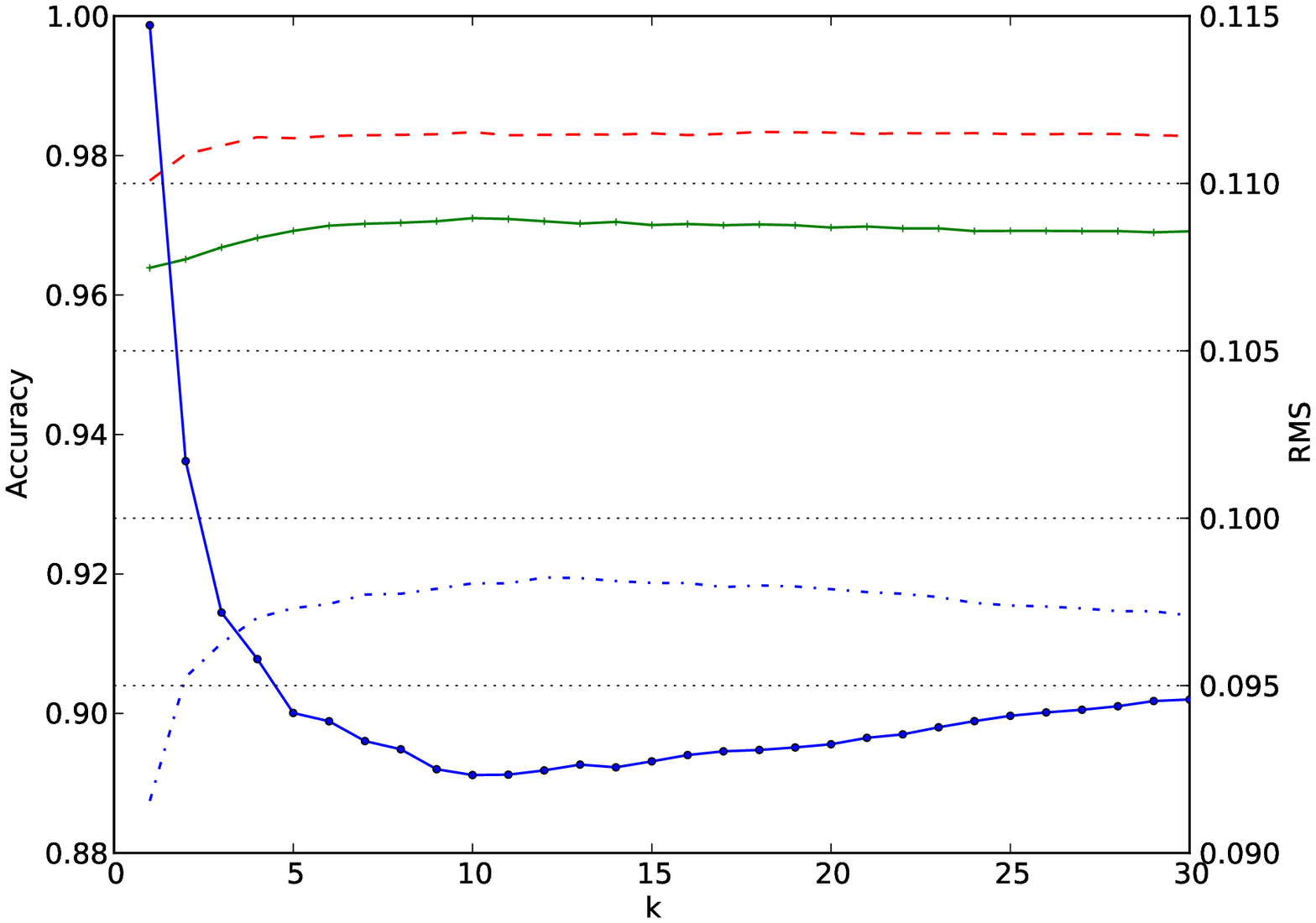}
\includegraphics[bb=28 180 594 586,width=6cm,clip]{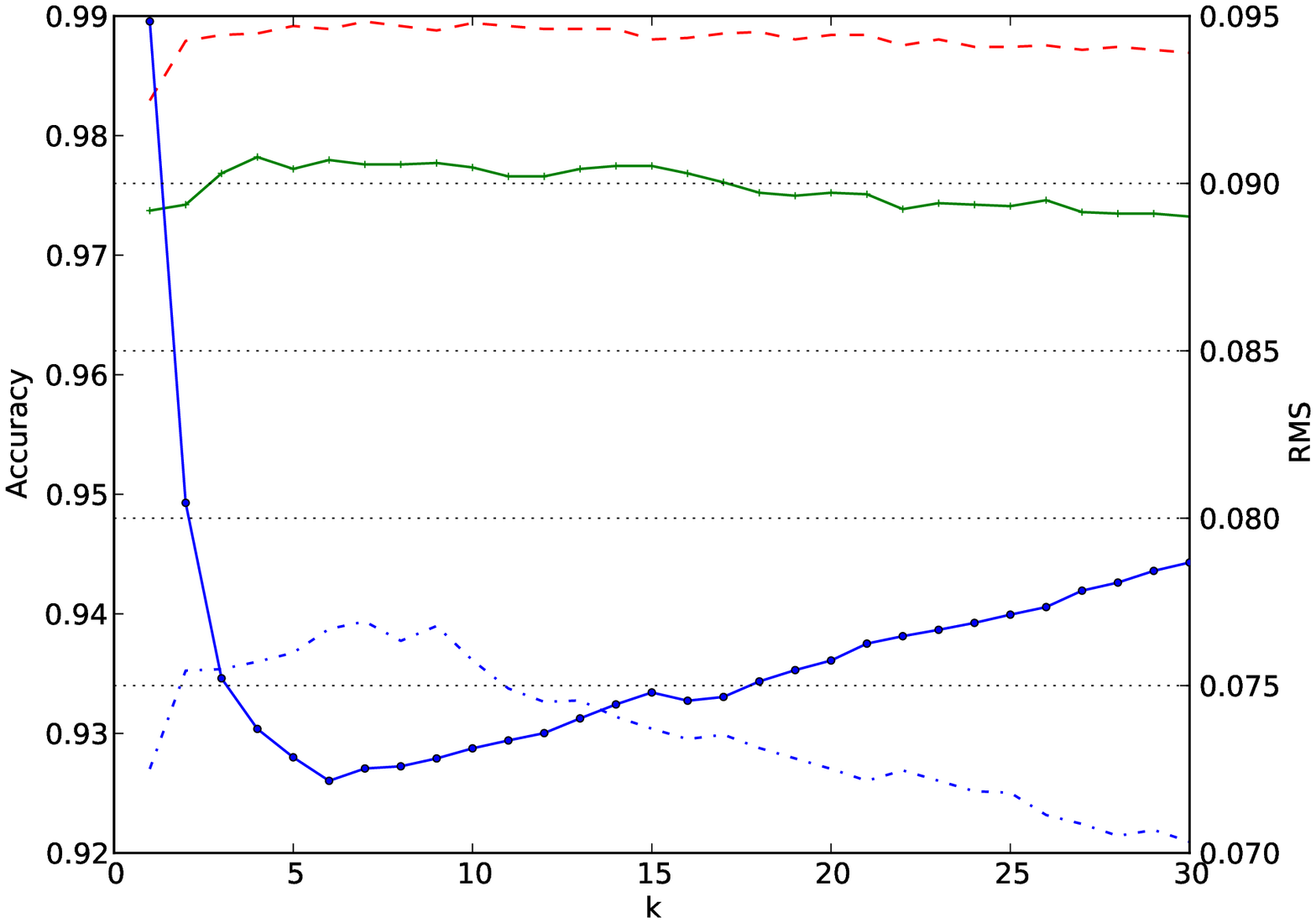}
\includegraphics[bb=28 180 594 586,width=6cm,clip]{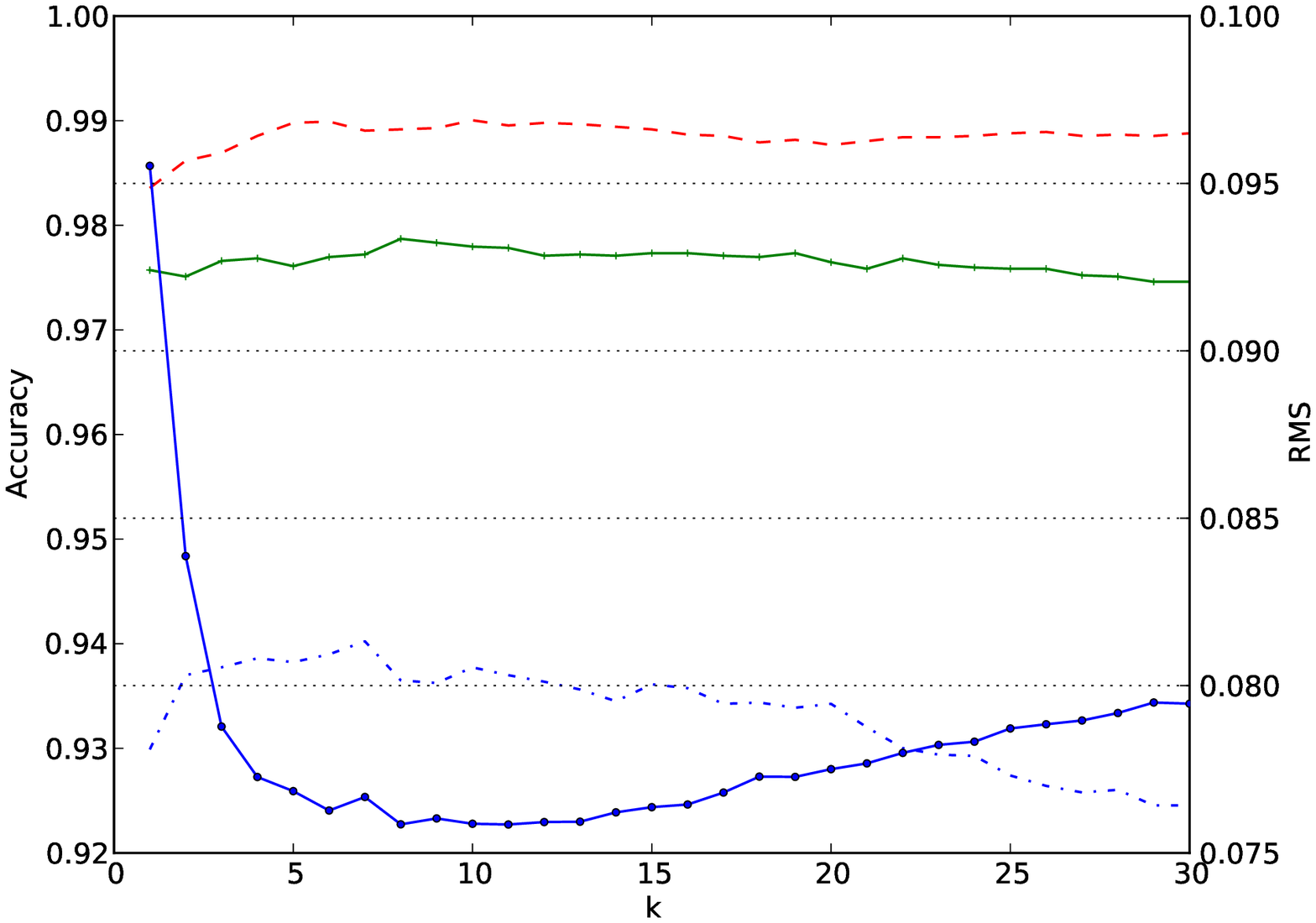}

  \caption{The accuracy within $|\Delta z| < 0.1$ (blue dot-dash-line),
$|\Delta z| < 0.2$ (green solid-line with plus) and $|\Delta z| <
0.3$ (red dash-line) and $RMS$ (blue solid-line with dot) as a
function of $k$ with the better input patterns for the SDSS sample,
the SDSS-UKIDSS sample, the SDSS-WISE sample and SDSS-UKIDSS-WISE
sample, respectively. For short, $4C = u-g,g-r,r-i,i-z$; $8C =
u-g,g-r,r-i,i-z,z-Y,Y-J,J-H,H-K$; $6C = u-g,g-r,r-i,i-z,z-W1,W1-W2$;
$10C = u-g, g-r, r-i, i-z, z-Y, Y-J, J-H, H-K, K-W1, W1-W2$. Left
panels: the input pattern is $4C+r$, $8C+r$, $6C+r$, $10C+r$,
respectively. Right panels: the input pattern is $4C+i$, $8C+i$,
$6C+i$, $10C+i$, respectively.}
\end{figure}

\begin{table}
\begin{center}
\caption{The value of $k$ neighbors for the better patterns for
different datasets. $4C = u-g,g-r,r-i,i-z$; $6C=
u-g,g-r,r-i,i-z,z-W1,W1-W2$; $8C = u-g,g-r,r-i,i-z,z-Y,Y-J,J-H,H-K$;
$10C = u-g, g-r, r-i, i-z, z-Y, Y-J, J-H, H-K, K-W1, W1-W2$.
}\label{tab:perf} {\tiny

\begin{tabular}{lcccccc}
\hline \hline
Data Set & Input Pattern &$k$&$|\Delta z| <0.1$ (\%) & $|\Delta z| <0.2$ (\%) & $|\Delta z| <0.3$ (\%) & $RMS$ \\
\hline
SDSS             & $4C,r$                            &17& 72.35$\pm$0.33 &83.94$\pm$0.21&89.87$\pm$0.30&0.201$\pm$0.003\\
SDSS             & $4C,i$                            &14& 72.50$\pm$0.29 &83.92$\pm$0.28&89.83$\pm$0.28&0.201$\pm$0.002\\
SDSS-UKIDSS      & $8C,r$                            &9 & 88.58$\pm$0.46 &94.90$\pm$0.51&96.88$\pm$0.20&0.115$\pm$0.006\\
SDSS-UKIDSS      & $8C,i$                            &7 & 89.17$\pm$0.37 &94.69$\pm$0.27&96.91$\pm$0.22&0.115$\pm$0.007\\
SDSS-WISE        & $6C,r$                            &9 & 91.58$\pm$0.30 &96.49$\pm$0.10&98.33$\pm$0.06&0.090$\pm$0.003\\
SDSS-WISE        & $6C,i$                            &10& 91.74$\pm$0.20 &97.02$\pm$0.07&98.31$\pm$0.06&0.093$\pm$0.003\\
SDSS-UKIDSS-WISE & $10C,r$                           &7 & 93.74$\pm$0.52 &97.63$\pm$0.21&98.95$\pm$0.25&0.083$\pm$ 0.010 \\
SDSS-UKIDSS-WISE & $10C,i$                           &5 & 93.82$\pm$0.44& 97.77$\pm$0.21&98.97$\pm$0.28&0.082$\pm$ 0.009\\

\hline
\end{tabular}}
\end{center}
\end{table}

In order to clearly show the difference among different samples, the
best predicted results for the four samples are presented in
Figure~3. For the SDSS sample, the best input pattern is $4C+r$ as
$k=17$; for the SDSS-UKIDSS sample, the best input pattern is $8C+i$
as $k=7$; for the SDSS-WISE sample, the best input pattern is $6C+r$
as $k=9$; and for the SDSS-UKIDSS-WISE sample, the best input
pattern is $10C+i$ as $k=5$. Comparisons of spectral redshifts with
estimated photometric redshifts are given in the left panel of
Figure~3, the distribution of $|\Delta z|$ is shown in the right
panel of Figure~3. It clearly shows that there are two clusters in
the first scatter plot of the SDSS sample, but they disappear
gradually by adding information from more bands, and when using data
from all bands in the three surveys, the number of discrete points
becomes very small. A small detail should be noticed that the points
near the diagonal line in the scatter plot of SDSS-WISE are thicker
than that of SDSS-UKIDSS. This indicates that adding WISE infrared
bands can help reduce the value of $RMS$ but raise the uncertainty
of photometric redshifts again. The influence of the number of bands
on the photometric redshift estimation is indicated in Figure~4. The
left panel of Figure~4 shows how $RMS$ of $|\Delta z|$ changes with
the number of photometric bands for the SDSS-UKIDSS sample (solid
line), the SDSS-WISE sample (dashed line) and the SDSS-UKIDSS-WISE
sample (dotted line), respectively. The right panel of Figure~4
indicates how percent outliers vary with the number of photometric
bands for the SDSS-UKIDSS sample (solid line), the SDSS-WISE sample
(dashed line) and the SDSS-UKIDSS-WISE sample (dotted line),
respectively. Outliers are defined as objects with $|\Delta
z|>0.15$. It is obvious that $RMS$ and percent outliers both usually
decrease when the number of photometric bands increases,
nevertheless, when the adding parameters have large uncertainty,
$RMS$ and percent outliers will rise. As a result, the performance
of a algorithm may not necessarily improve when adding more
parameters from more bands to predict photometric redshifts.

\begin{figure}[!!!h]
\includegraphics[bb=39 128 563 651,width=5.5cm,clip]{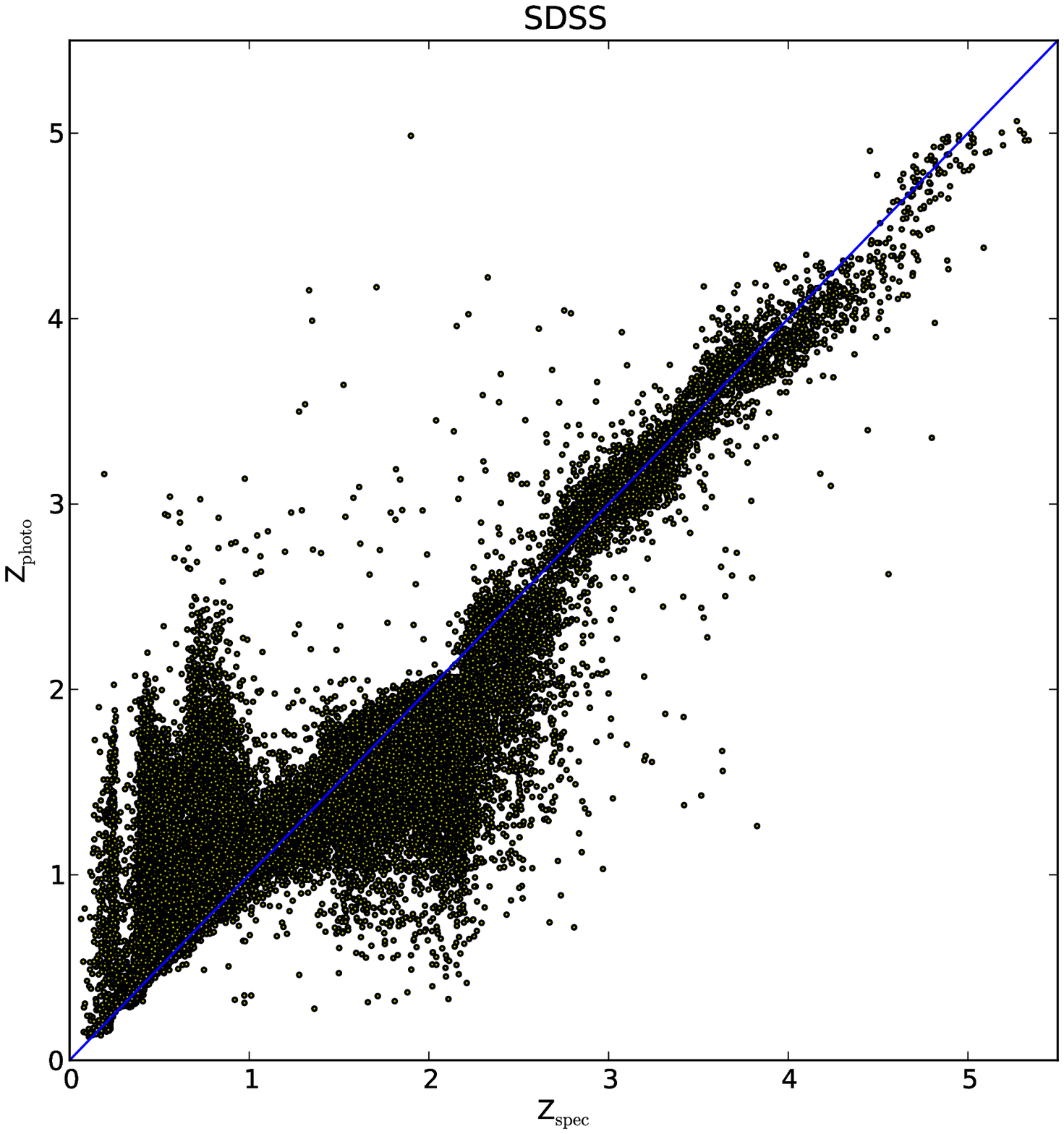}
\includegraphics[bb=39 192 448 589,width=5.5cm,clip]{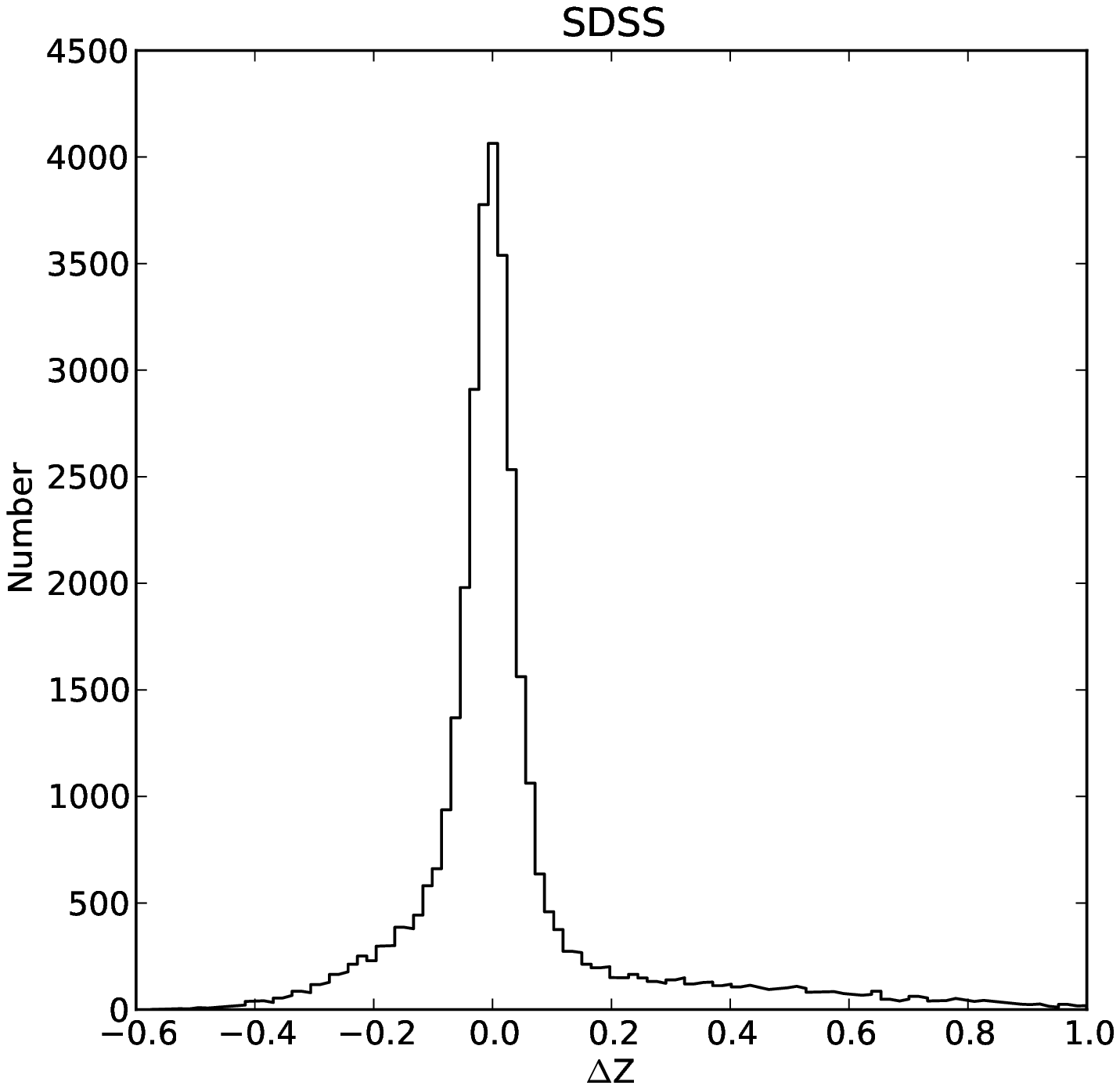}
\includegraphics[bb=39 128 563 651,width=5.5cm,clip]{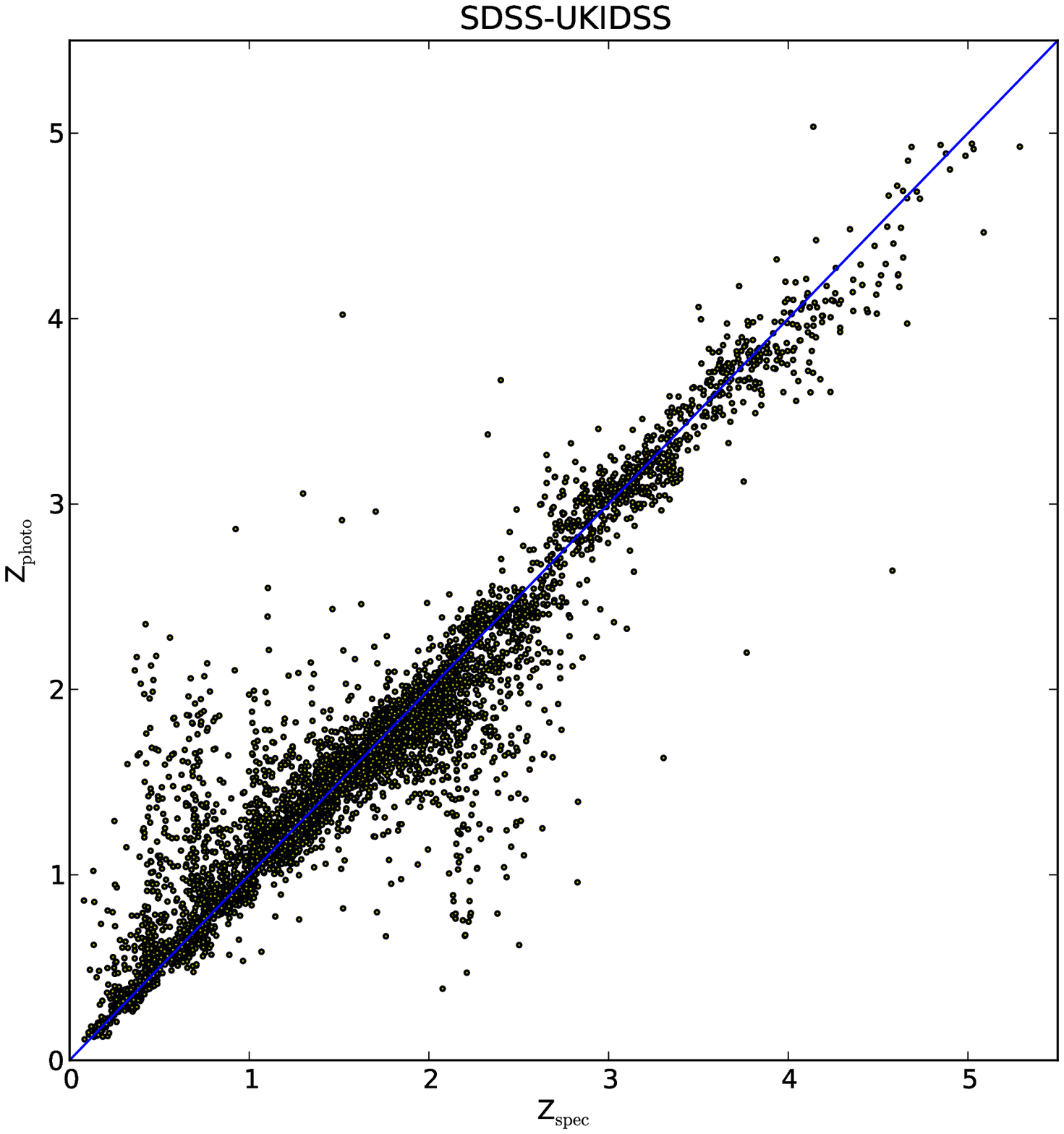}
\includegraphics[bb=39 192 448 589,width=5.5cm,clip]{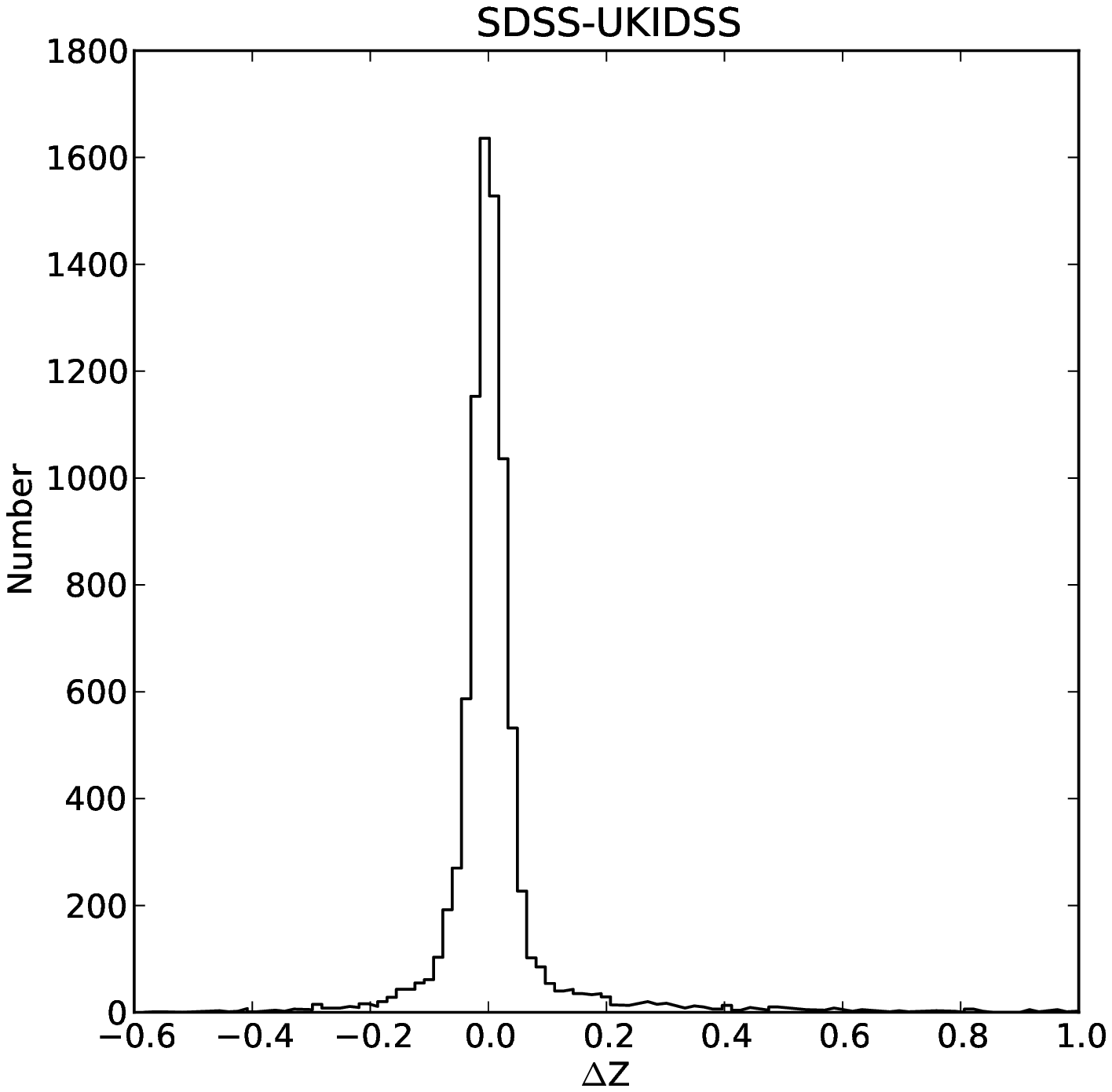}
\includegraphics[bb=39 128 563 651,width=5.5cm,clip]{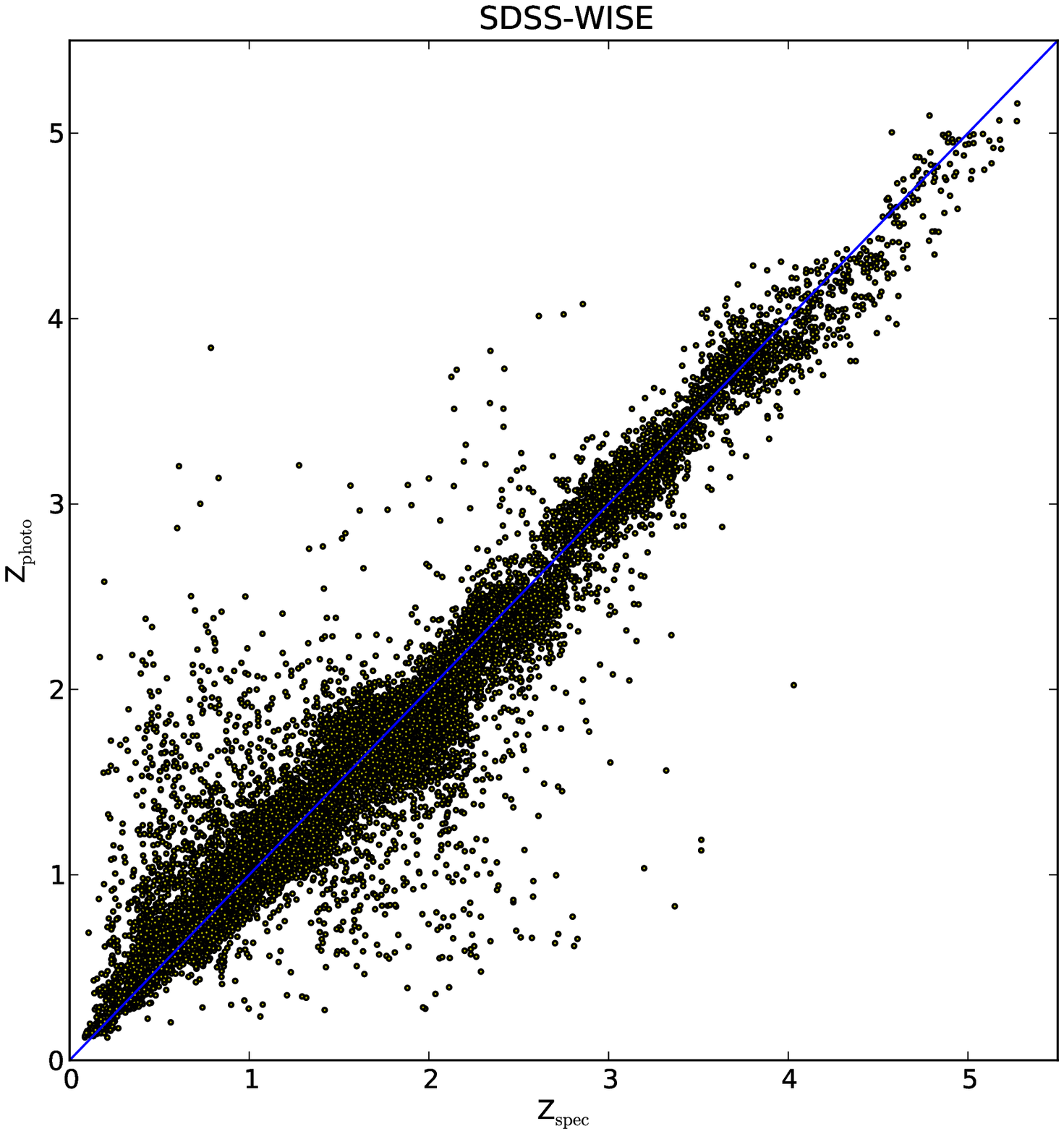}
\includegraphics[bb=39 192 448 589,width=5.5cm,clip]{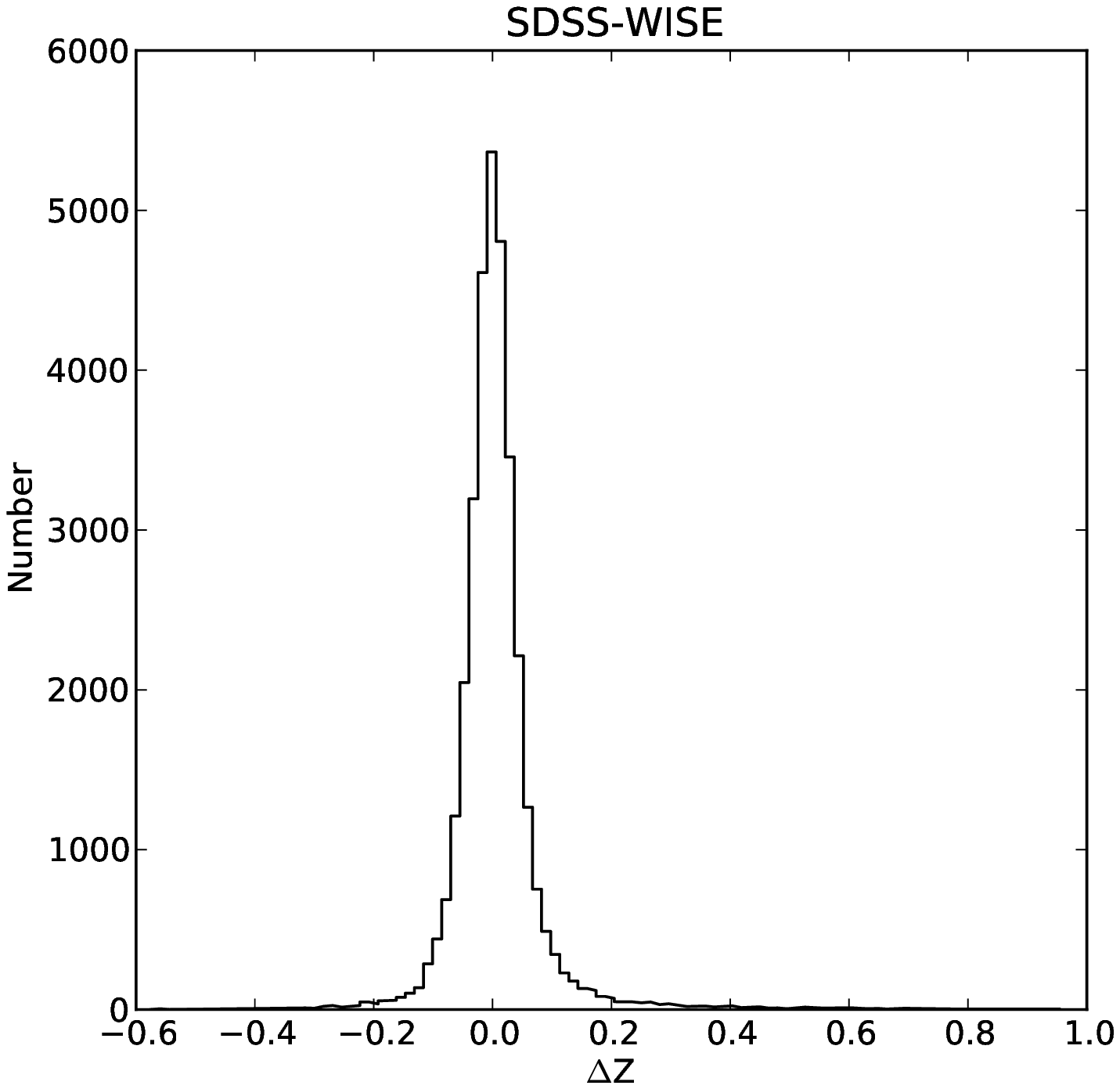}
\includegraphics[bb=39 128 563 651,width=5.5cm,clip]{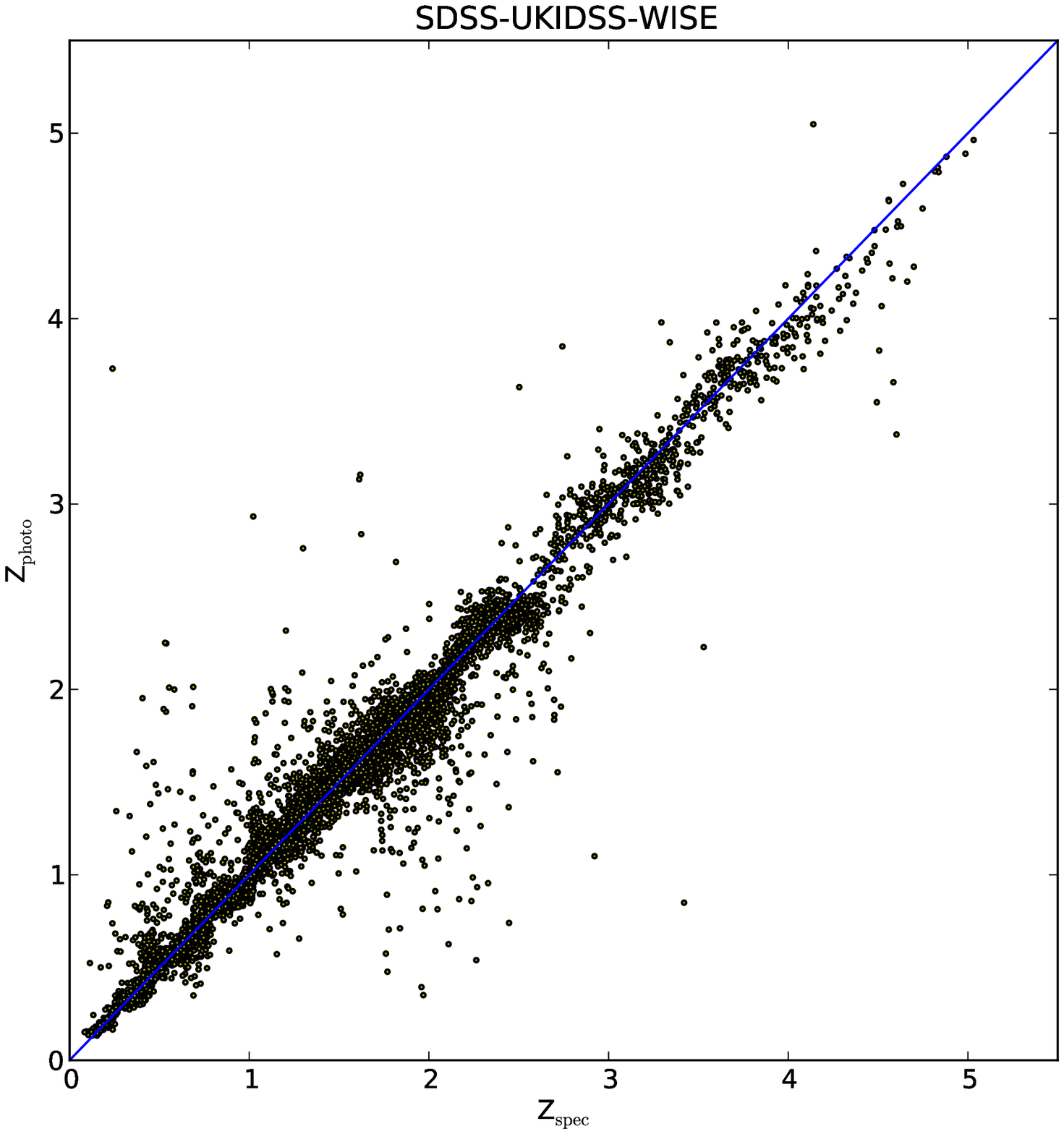}
\includegraphics[bb=39 192 448 589,width=5.5cm,clip]{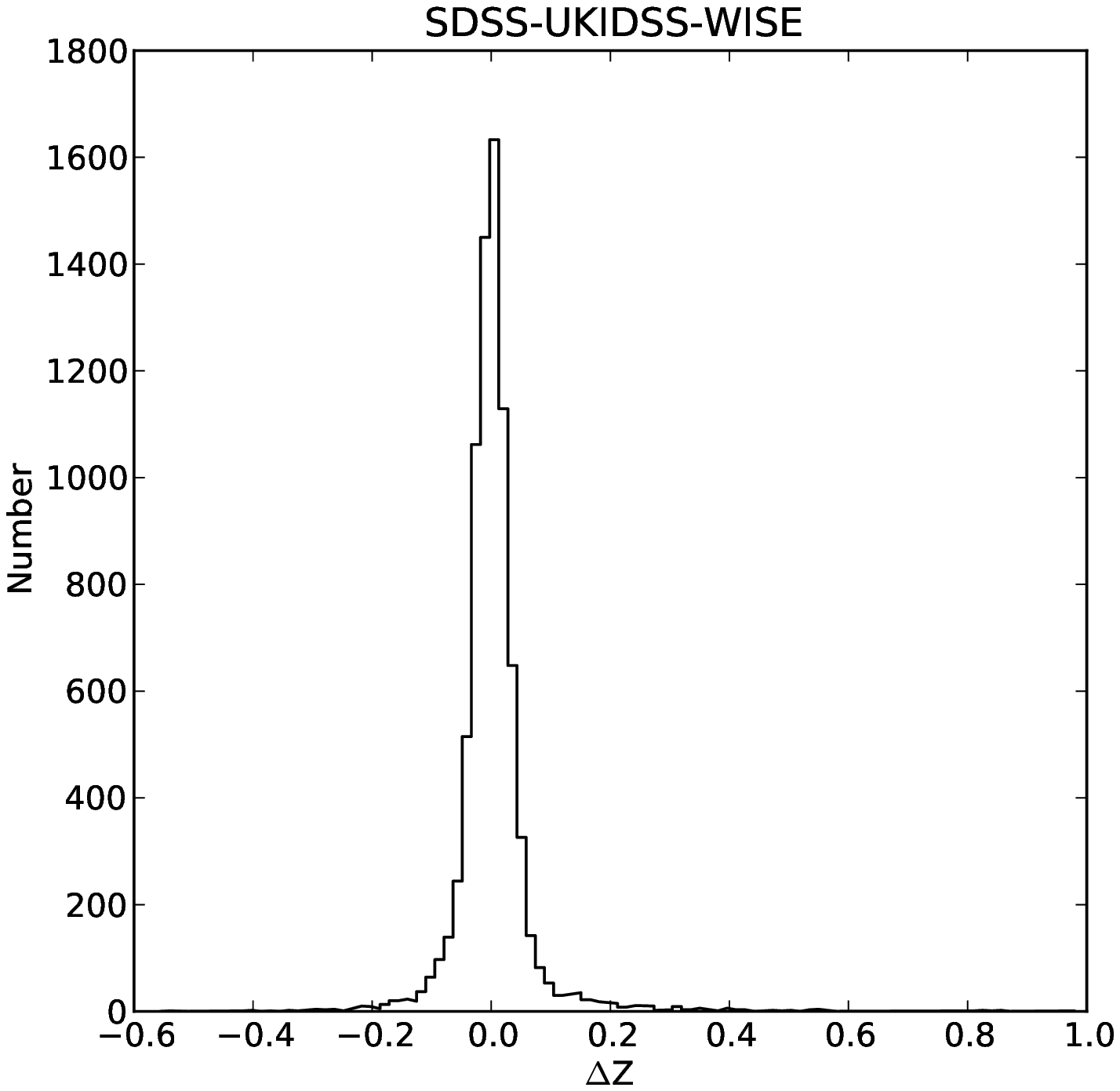}

  \caption{Comparison of photometric redshift estimation results for the SDSS sample with the
best input pattern of $4C+r$ as $k=17$; for the SDSS-UKIDSS sample
with the best input pattern of $8C+i$ as $k=7$, for the SDSS-WISE
sample with the best input pattern of $6C+r$ as $k=9$; and for the
SDSS-UKIDSS-WISE sample with the best input pattern of $10C+i$ as
$k=5$.}
\end{figure}

\begin{figure}
\includegraphics[bb=31 31 502 502,width=8cm,clip]{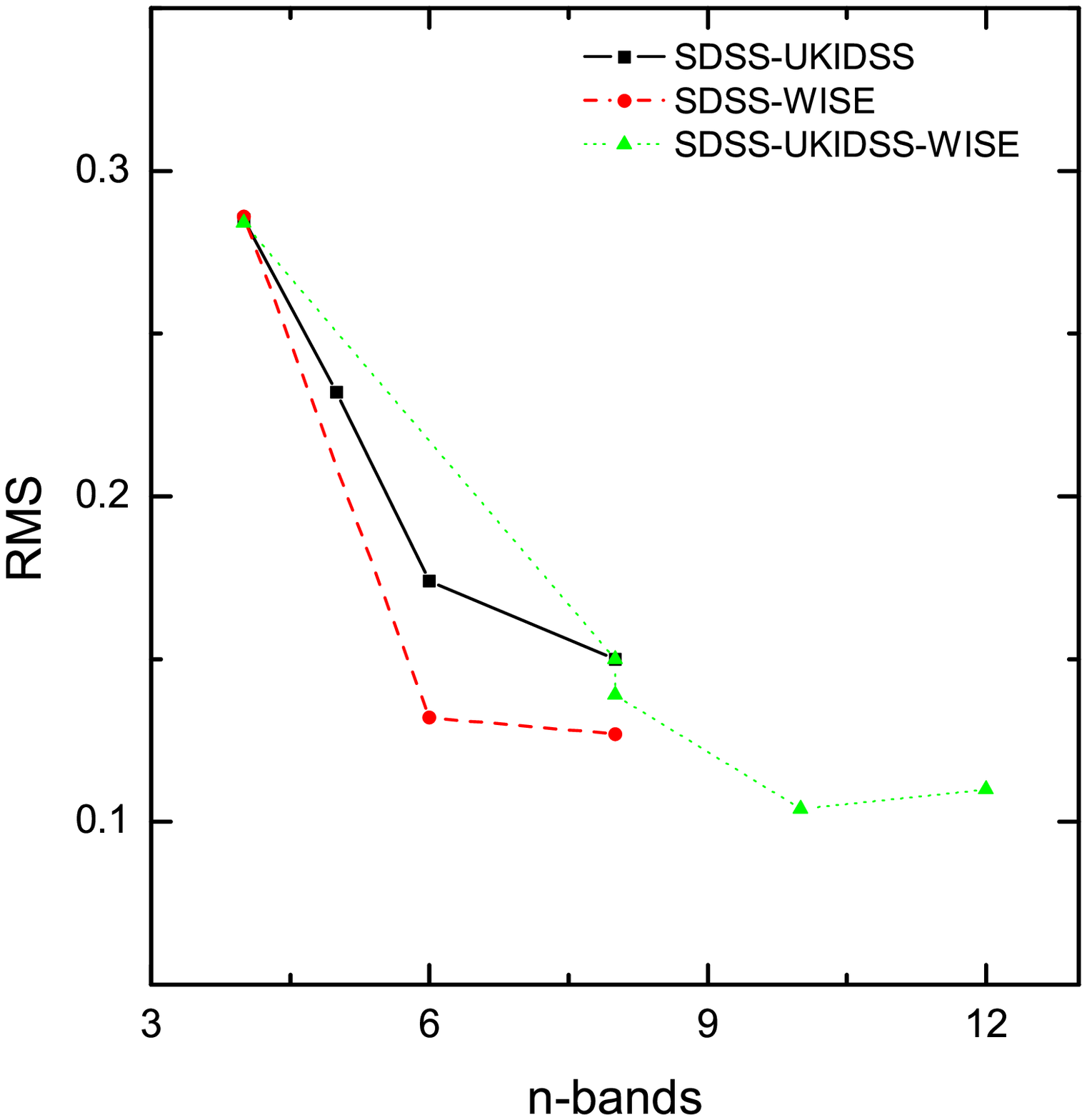}
\includegraphics[bb=31 31 502 502,width=8cm,clip]{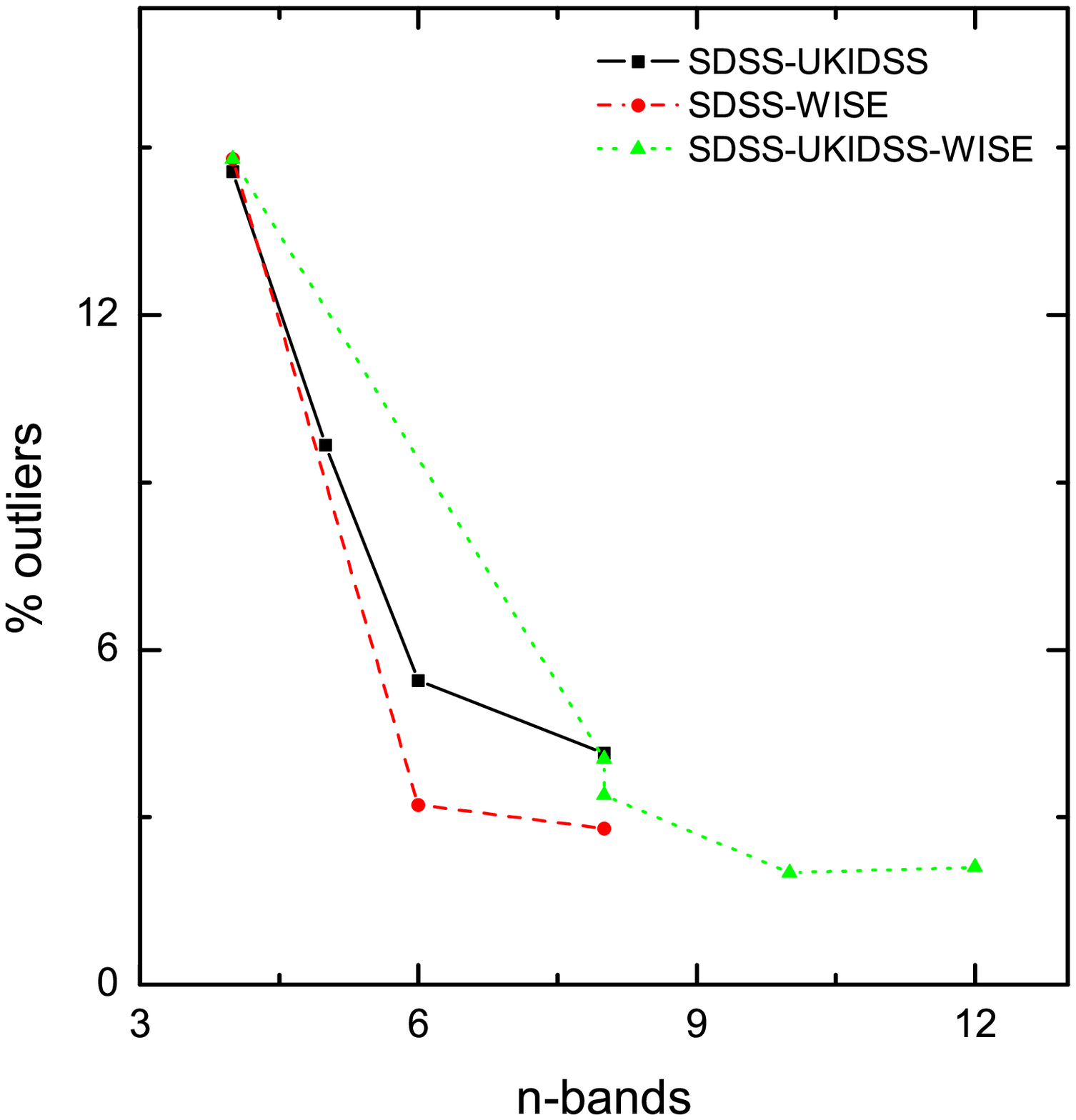}

\caption{Left panel: $RMS$ of $(z_{\rm phot}-z_{\rm spec})/(1 +
z_{\rm spec})$ versus the number of photometric bands for the
SDSS-UKIDSS sample (solid line), the SDSS-WISE sample (dashed line)
and the SDSS-UKIDSS-WISE sample (dotted line), respectively. Right
panel: percent outliers versus the number of photometric bands for
the SDSS-UKIDSS sample (solid line), the SDSS-WISE sample (dashed
line) and the SDSS-UKIDSS-WISE sample (dotted line), respectively.}
\end{figure}

\subsection{Comparison with other algorithms}

Various algorithms have been applied on photometric redshift
estimation of quasars. The performance of algorithms depends on many
factors, such as algorithm itself, model parameters, data, input
pattern. It is difficult to say which algorithm is the best. We only
think the method is the best for the special data when its model
parameters and input pattern are set. For giving a rough comparison
of algorithms on this issue, all models we study in this subsection
are from WEKA (The Waikato Environment for Knowledge Analysis). WEKA
is a tool for data analysis and includes implementations of data
pre-processing, classification, regression, clustering, association
rules, and visualization by different algorithms. All these
algorithms can refer to the book published by Witten \& Frank
(2005). WEKA's binaries and sources may be freely available from
http://www.cs.waikato.ac.nz/ml/weka/. For our case, the function
algorithms implemented include linear regression, Multi-layer
perceptron (MLP), pace regression, Radial Basis Function (RBF)
network, Support Vector Machines (SVM); lazy learning methods used
consist of IB1, IB5 and Locally Weighted Learning (LWL), here IB1
and IB5 all belong to $k$NN, IB1 corresponds to $k$NN when $k=1$
while IB5 points to $k$NN as $k=5$; treelike algorithms applied are
REPTree and Decision Stump; so-called ``meta" model utilized is
bagging; rule models used contain M5-Rules Algorithm and decision
table.

The comparison of all models is made on the basis of the following
criteria:

i) Correlation Coefficient (also known as $R$): $R$ is a measure of
the strength and direction of the linear relationship between two
variables that is defined in terms of the (sample) covariance of the
variables divided by their (sample) standard deviations.

i) Mean absolute error ($MAE$): MAE is the average of the difference
between predicted and actual value in all test cases.

iii) Root Mean-Squared Error ($RMSE$): $RMSE$ is commonly used
measure of differences between predicted values (photometric
redshifts) by a model and the actually observed values (spectral
redshifts). The mean-squared error is one of the most frequently
used measures of success for numeric prediction. This value is
obtained by computing the average of the squared differences between
each computed value and its corresponding correct value. $RMSE$ is
just the square root of the mean-squared-error.

Correlation Coefficient ($R$), mean absolute error ($MAE$) and root
mean squared error ($RMSE$) is calculated for each machine learning
algorithm, respectively. For brief, the input pattern $10C+i$ is
taken for all the models. All the experiments are done by 10-fold
cross-validation. The comparison of the performance of different
algorithms for predicting photometric redshifts of quasars are shown
in Table~3. The time to build a model is also depicted in Table~3.
As Table~3 indicated, MLP of all function models shows the best
performance with the largest value of $R$ and the smallest values of
$MAE$ and $RMSE$; IB5 (i.e $k$NN with $k=5$) of lazy learning
methods is the best with $R=0.973, MAE=0.096$ and $RMSE=0.186$;
REPTree of treelike algorithms is superior to decision stump;
M5-rules of rule methods outperforms decision table. Bagging is a
popular method that improves the performance for any learning
algorithm. The performance of REPTree with bagging algorithm indeed
improves compared to REPTree as Table~3 described. Taking no account
of lazy learning, the fastest speed to create a training model is
pace regression, the second is decision stump, the third belongs to
linear regression, however, the three approaches have no good
performance for photometric redshift prediction; the slowest is SVM.
Considering all algorithms applied here, IB5 shows the best
performance. This is why we utilize $k$NN for our case. The
advantage of lazy learning is that the training and prediction all
delay until a query is made to the system, it deals successfully
with changes in the problem domain and it is very useful for large
data with few attributes; the disadvantages of lazy learning require
the large space to store the entire training data, another
disadvantage is that lazy learning approaches are usually slower to
predict, though they have a faster training phase. Nevertheless,
$k$NN based on KD-Tree improves the speed to evaluate, just like
that is done in this paper.

\begin{table}
\begin{center}
\caption{Comparison of various methods for photometric redshift
estimation of quasars with the input pattern $10C+i$. } {

\begin{tabular}{lccccc}
\hline \hline
Class&Method &$R$ &$MAE$& $RMSE$&Time(s) \\
\hline
function&Linear Regression    &0.887&0.274&0.369&0.21\\
&MLP&0.910&0.256&0.360&37.92 \\
&Pace Regression      &0.887&0.274&0.369&0.06 \\
&RBF Network&0.395&0.568&0.734&3.26\\
&SVM&0.884&0.272&0.373&606.61  \\
\hline
lazy learning&IB1      &0.959&0.112&0.228&0\\
&IB5      &0.973&0.096&0.186&0\\
&LWL&0.665&0.472&0.567&0\\
\hline
tree&REPTree               &0.936&0.140&0.282&0.45\\
&Decision Stump        &0.627&0.498&0.622&0.1\\
\hline
meta&Bagging of REPTree      &0.962&0.115&0.219&3.05\\
\hline
rule&Decision Table        &0.861&0.260&0.406&3.89\\
&M5-Rules               &0.956&0.128&0.235&59.9\\

\hline
\end{tabular}}
\end{center}
\end{table}

\section{CONCLUSIONS}
\label{sec:conclusion}

We have demonstrated a simple and useful method $k$-nearest neighbor
using multiband data which can effectively estimate photometric
redshifts of quasars. According to the above experiments and the
comparison with other algorithms, $k$NN can be taken as a powerful
tool to determine photometric redshifts of quasars, when it use
$k=5$ and ($u-g, g-r, r-i, i-z, z-Y, Y-J, J-H, H-K, K-W1, W1-W2, i$)
as the input pattern. On the SDSS-UKIDSS-WISE data set, the accuracy
within $|\Delta z|<0.1$ can reaches $93.82\% \pm 0.44\%$, $97.77\%
\pm 0.21\%$ within $|\Delta z|<0.2$, $98.97\% \pm 0.28\% $ within
$|\Delta z|<0.3$ and $RMS$ is $0.082 \pm 0.009$. This method
successfully avoids the redshift catastrophic failure when the
genuine redshift of quasar is below 2.8. Actually, $k$NN can reach
the same level of performance only using cross-match data from the
SDSS and UKIDSS databases, but $RMS$ is a little higher (0.115).
Adding information from the WISE infrared band can effectively
improve the performance of $k$NN on $RMS$. Finally, we improve the
performance of $k$NN within $|\Delta z|<0.3$ from 85.19\% to about
98.97\% and reduce $RMS$ from 0.287 to 0.009. This means that $k$NN
using multiband data is an effective and feasible approach for
estimating photometric redshifts of quasars. The major advantages of
$k$NN are its simplicity and easy understanding without physical
assumption. It does not require training, takes the known
spectroscopic redshifts as templates and further shows its
superiority with large numbers of spectra available. Moreover, the
$k$NN approach combined with KD-Tree conquers the high
time-consumption faced by traditional $k$NN method.

In the future, the main target is about how to improve the accuracy
of photometric redshifts on $|\Delta z|<0.1$. Without doubt, using
multibands is a useful and reliable way to improve it and we plan to
add ultraviolet band data from GALEX to test this method. Another
possible solution is about that $k$NN can be combined with some
classification method to conquer the catastrophic failure area on
the SDSS optical band that is met by many machine learning
algorithms. The work of photometric redshift estimation on optical
band is important and urgent because the overlapped area among the
several surveys are relatively small and there is over billion of
unknown objects with photometric data. We will implement it for
predicting the photometric redshifts of quasar candidates for the
Guoshoujing Telescope (LAMOST) or other projects in the world.

\section*{Acknowledgments}
We are very grateful to the referee for his/her valuable comments
and suggestions, which helped improve the quality of our paper
significantly. This paper is funded by National Natural Science
Foundation of China under grant No.10778724, 11178021 and
No.11033001, the Natural Science Foundation of Education Department
of Hebei Province under grant No. ZD2010127. This work has made use
of data products from the Sloan Digital Sky Survey (SDSS), the UKIRT
Infrared Deep Sky Survey (UKIDSS) and Wide-field Infrared Survey
Explorer (WISE).


\begin{thebibliography}{}
\bibitem[{()}]{} Abdalla, F. B., Banerji, M., Lahav, O., Rashkov, V., 2011,
MNRAS, 417, 1891
\bibitem[{()}]{} Abazajian, K. N. et~al. 2009, ApJS, 182, 543
\bibitem[{()}]{} Babbedge, T. S. R., Rowan-Robinson, M., Gonzalez-Solares, E.,
et~al. 2004, MNRAS, 353, 654
\bibitem[{()}]{} Ball, N. M. \& Brunner, R. J. 2010, IJMPD, 19, 1049
\bibitem[{()}]{} Ball, N. M., Brunner, R. J., Myers, A. D., Strand, N. E., Alberts, S. L., Tcheng, D., Llora, X. 2007, ApJ, 663, 774
\bibitem[{()}]{} Ball, N. M., Brunner, R. J., Myers, A. D., Strand, N. E., Alberts, S. L., Tcheng, D. 2008, ApJ, 683, 12
\bibitem[{()}]{} Bolzonella, M., Miralles, J.-M., Pell{\'o} R. A\&A, 2000, 363, 476
\bibitem[{()}]{} Bovy, J. et al. 2012, ApJ, 749, 41
\bibitem[{()}]{} Blake, C. et~al. 2007, MNRAS, 374, 1527
\bibitem[{()}]{} Borne, K., in Next Generation of Data Mining (Taylor \& Francis: CRC Press), pp. 91-114 (2009)
\bibitem[{()}]{} Budav$\acute{a}$ri, T., Csabai, I., Szalay, A. S., et~al., 2001, AJ, 122,
1163
\bibitem[{()}]{} Budav$\acute{a}$ri, T., Szalay, A. S., Connolly, A. J., Csabai, I., Dickinson, M. 2000,
AJ, 120, 1588
\bibitem[{()}]{} Collister, A. A. \& Lahav, O., 2004, PASP, 116, 345
\bibitem[{()}]{} Firth, A. E., Lahav, O., Somerville, R. S., 2003, MNRAS, 349, 1397
\bibitem[{()}]{} Hewett, P. C., Warren, S. J., Croom, S. M. 2008, MNRAS, 365, 1605
\bibitem[{()}]{} Hall, M., Frank, E., Holmes, G., Pfahringer, B., Reutemann, P., Witten, I. H. 2009, SIGKDD Explorations, 11(1), 10
\bibitem[{()}]{} Hildebrandt, H., Arnouts, S., Capak, P., et~al., 2010, A\&A,
523, A31
\bibitem[{()}]{} Geach, J. E. 2011, MNRAS, 419, 2633
\bibitem[{()}]{} Kaiser, N. \& Aussel, H. 2002, \spie, 4836, 154
\bibitem[{()}]{} Kumar, N. D. 2008, Mater Dissertation, St Anne's
College, University of Oxford
\bibitem[{()}]{} Lawrence, A. et~al. 2007, \mnras, 379, 1599
\bibitem[{()}]{} Peng, N., Zhang, Y., Zhao, Y., 2010a, Proc. SPIE, 7740, 92
\bibitem[{()}]{} Peng, N., Zhang, Y., Zhao, Y., 2010b, Proc. SPIE, 7740, 86
\bibitem[{()}]{} Polsterer, K. L., Zinn, P.-C., Gieseke, F. 2013, MNRAS,
428, 226
\bibitem[{()}]{} Oyaizu, H., Lima, M., Cunha, C. E., Lin, H., Frieman, J., 2008, ApJ, 674, 768
\bibitem[{()}]{} Tyson, J. A. 2002, \spie, 4836, 10
\bibitem[{()}]{} Wadadekar, Y., et~al. 2005, PASP, 117,79
\bibitem[{()}]{} Wang, D., Zhang, Y., Liu, C., Zhao, Y. 2008, ChJAA, 8(1), 119
\bibitem[{()}]{} Witten, I. H., Frank, E., Data Mining: Practical Machine Learning
Tools and Techniques with Java Implementations, Morgan Kaufmann, San
Francisco, 2005
\bibitem[{()}]{} Laurino, O., D'Abrusco, R., Longo, G., Riccio, G. 2011, MNRAS, 418, 2165
\bibitem[{()}]{} Warren, S. J., Hewett, P.C., Foltz, C.B. 2000, MNRAS, 312, 827
\bibitem[{()}]{} Way, M. J., Foster, L. V., Gazis, P. R., Srivastava, A. N. 2009, ApJ, 706, 623
\bibitem[{()}]{} Richards G. T., et al. 2001, AJ, 122, 1151
\bibitem[{()}]{} Salvato, M., Hasinger, G., Ilbert, O., et~al. 2009, ApJ, 690, 1250
\bibitem[{()}]{} Salvato, M., Ilbert, O., Hasinger, G., et~al. 2011, ApJ,
742, 61
\bibitem[{()}]{} Schlegel D. J., Finkbeiner D. P., Davis M., 1998, ApJ,
500, 525
\bibitem[{()}]{} Schneider, D. P., et al. 2010, AJ, 139, 2360
\bibitem[{()}]{} Weinstein, M. A., et al. 2004, ApJS, 155,243
\bibitem[{()}]{} Wright, E.L. et al. 2010, AJ, 140, 1868
\bibitem[{()}]{} Wu, X.-B., Zhang, W., Zhou, X. 2004, ChJAA, 4, 17
\bibitem[{()}]{} Warren, S. J., Hewett, P. C., Foltz, C. B. 2000, MNRAS, 312,827
\bibitem[{()}]{} Wu, X.-B. \& Jia, Z.-D., 2010,
\mnras, 406, 1583
\bibitem[{()}]{} Y\'eche, C. et~al. 2010, A\&A, 523, A14
\bibitem[{()}]{} York, D. G., et~al. 2000, AJ, 120, 1579
\bibitem[{()}]{} Zhang, Y., Li, L., Zhao, Y. 2009, MNRAS,
392,233
\bibitem[{()}]{} Zhang, Y., Zheng, H., Zhao, Y. 2008, Proc. SPIE, 7019,
701938-1


\end{thebibliography}
\end{document}